\documentclass[aps,  pra,  superscriptaddress,  showpacs]{revtex4-1}
\usepackage[T1]{fontenc}
\usepackage[utf8]{inputenc}
\usepackage[english]{babel}
\usepackage{amsmath,amssymb,amsfonts}
\usepackage{graphicx,graphics}
\usepackage{subfig}
 \usepackage{caption}
 \usepackage{dcolumn}
\usepackage{bm}
\usepackage{color}
\usepackage{amsthm}             %para el teorem style
\usepackage{amssymb}
\usepackage{mathrsfs}
\usepackage{dcolumn}% Align table columns on decimal point
\usepackage{bm}% bold math
\usepackage{epsfig}
\usepackage{epstopdf}
\usepackage{float}

\begin{document}

\title{Photons in the presence of  parabolic mirrors }
\author{R. Guti\'errez-J\'auregui}
\affiliation{The Dodd-Walls Centre for Photonic and Quantum Technologies, Department of Physics, University of Auckland,
Private Bag 92019, Auckland, New Zealand}
\author{R. J\'auregui}
\affiliation{Instituto de F\'{\i}sica, Universidad Nacional Aut\'onoma de M\'exico, Apdo. Postal 20-364, 01000 Cd. de M\'exico, M\'exico}
\email{rocio@fisica.unam.mx}
\begin{abstract}
We present a vectorial analysis of the behavior of the electromagnetic field in the presence of boundaries with parabolic geometry.
The relevance of the use of symmetries to find explicit closed expressions for the electromagnetic fields is emphasized.
Polarization and phase related  angular momenta of light have an essential role in the proper definition of the generator $\mathfrak{A}_3$ 
 of a symmetry transformation  that distinguishes the parabolic geometry. 
Quantization of the electromagnetic field in terms of the resulting elementary modes is performed.
The important case of a boundary defined by an ideal parabolic mirror is explicitly worked out. The presence of the mirror restricts the eigenvalues 
of $\mathfrak{A}_3$ available to the electric  and magnetic   fields of a given mode, via compact expressions. Modes previously reported in the literature are
particular cases of those described in this work.     
\end{abstract}

%\pacs{42.50.-p, 42.65.Lm, 32.80.Lg}
\maketitle

\section{Introduction}

Due to its focusing properties,  the parabola of revolution has been considered an optimal geometry to  build mirrors and lenses since classical antiquity.  Parabolic mirrors are usually designed under conditions that allow a ray description of the relevant electromagnetic (EM) modes. Recently, there have been attempts to make a detailed description of the vectorial classical \cite{noeckel} and quantum \cite{luis} properties of the electromagnetic field in the presence of parabolic boundaries; a motivation being  the possibility of optimizing the coupling of single atoms to single photons. The basic idea behind this optimization is that the processes of elastic scattering \cite{scattering,leong} and absorption \cite{absorption, paul} of a photon by a single atom increase their efficiencies for incident light that spatially resembles the natural mode of the atomic transition, which usually corresponds to an electric dipole wave. It has been shown that a deep parabolic mirror can  focus a radially polarized doughnut mode to a field that is nearly linearly polarized along the optical axis and, close to the focus of the parabola,  is similar to the dipole field \cite{donut,Sonder,dorn}. Experiments working on this direction have lead to a better understanding of the interaction between photons and atomic systems under controlled conditions with remarkable results \cite{pinosti,wilk,experiments}.

Previous analysis of the EM field  in the presence of parabolic boundaries, however,  have been limited by mathematical difficulties in finding a complete set of elementary modes that fulfill both Maxwell equations and the adequate boundary conditions. As a consequence, important properties like space dependencies of the polarization are not fully understood. Nevertheless, these works have shown that
particular EM modes in the presence of ideal parabolic mirrors would exhibit several interesting properties. For instance, a WKB study of the dynamics within a parabolic cavity indicate that the waves without optical vortices should be robust with respect to small geometrical deformations of the cavity, while those exhibiting optical vortices could be unstable and even give rise to optical chaos \cite{noeckel}.

In this work, we present a detailed description of the electromagnetic vectorial field in parabolic geometries that surpasses the problems mentioned above. We construct a complete set of modes that satisfy Maxwell equations and incorporate the underlying symmetries explicitly. These modes could allow a clearer description of experimental arrays involving parabolic boundaries, since they can be used to explore properties that cannot be accessed through previous descriptions. We also make a proper quantization of the EM field in terms of the classical Maxwell modes; this provides an extended framework for the studies mentioned above regarding the interaction between light and matter under controlled conditions. We illustrate the relevance of this approach by working out in detail the paradigmatic configuration of the EM field  in the presence of an ideal parabolic mirror.

In the next section we revisit the characteristics of scalar waves in parabolic coordinates. Following Boyer {\it et al} \cite{boyer}, the generators of the natural symmetries are identified, as well as the angular spectrum of the scalar modes. In Section III, we obtain the solutions of the Maxwell equations. The EM modes are written in terms of vectorial Hertz modes $\boldsymbol{\pi}$ with components that
are, in turn, written in terms of the scalar modes. The $\boldsymbol{\pi}$ modes are chosen to yield electric $\mathbf{E}$ and magnetic $\mathbf{B}$ fields that are eigenvectors of the generators of the mentioned symmetries with the same eigenvalues; a key point is that the generators of the adequate transformations  act now  on the {\it vector} EM fields. Intrinsic and orbital angular momenta of light have an essential role in the proper definition of the generator $\hat{\mathfrak{A}}_3$ 
 of a symmetry transformation  that distinguishes the parabolic geometry. 
The presence of an ideal parabolic mirror restricts the eigenvalues of $\hat{\mathfrak{A}}_3$ available to the $\mathbf{E}$ and $\mathbf{B}$ of a given mode, via compact expressions. The modes that were worked out before are contained within the set of modes we get in this manuscript. The elementary modes we find are orthogonal and can be normalized according to Einstein prescription. In this way the quantum EM field  operators are constructed.

\section{Scalar parabolic modes}
The parabolic coordinates are defined by:
\begin{equation}
x_1=\sqrt{\zeta\eta}\cos\varphi,\quad\quad x_2=\sqrt{\zeta\eta}\sin\varphi,\quad\quad x_3=\frac{1}{2}(\zeta - \eta)
\end{equation}
with
\begin{equation}
0\le\zeta<\infty,\quad\quad 0\le\eta<\infty,\quad\quad 0\le\varphi<2\pi.
\end{equation}
In Appendix A, explicit expressions for the scale factors $h_\zeta$, $h_\eta$ and $h_\varphi$, as well as the unitary parabolic vectors $\mathbf{e}_\zeta$, $\mathbf{e}_\eta$ and $\mathbf{e}_\varphi$ in terms of the Cartesian basis $\{\mathbf{e}_1, \mathbf{e}_2,\mathbf{e}_3 \}$ are given.
\begin{figure}
\includegraphics[scale=1.25]{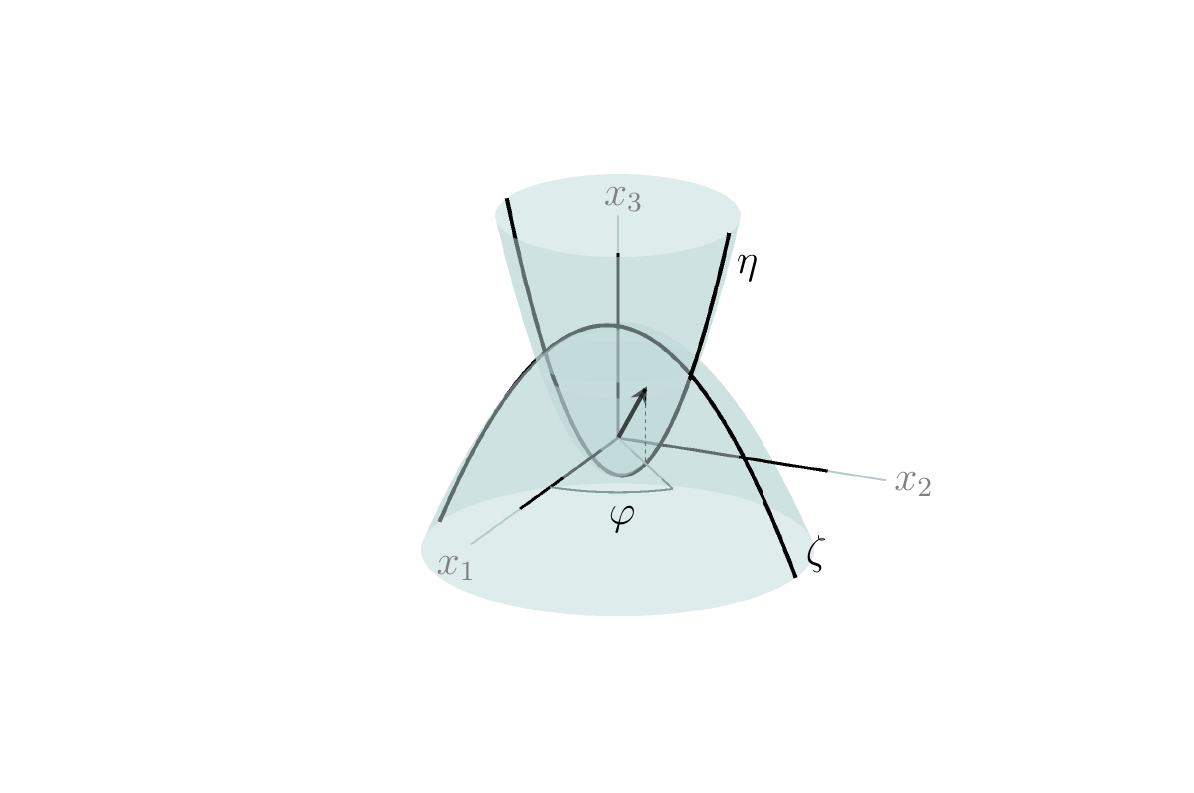}
\caption{Parabolic coordinates $\{ \zeta, \eta, \varphi\}$. Surfaces of constant $\zeta$ ($\eta$) correspond to downward (upward) paraboloids of revolution about the $x_3$-axis. The foci of all parabolas is located at the origin. The third coordinate $\varphi$ corresponds to the azimuth angle.}\label{fig:coor}
\end{figure}

The wave equation in parabolic coordinates
\begin{eqnarray}
\boldsymbol{\nabla}^2\Psi(\zeta,\eta,\varphi) &=& \frac{\omega^2}{c^2}\Psi(\zeta,\eta,\varphi)\\
&=& \frac{4}{\zeta + \eta} \Big[\frac{\partial}{\partial\zeta}\zeta \frac{\partial}{\partial\zeta} +\frac{\partial}{\partial\eta}\eta \frac{\partial}{\partial\eta}\Big]\Psi +\frac{1}{\zeta\eta}\frac{\partial^2\Psi}{\partial\varphi^2}
\end{eqnarray}
is separable, so that the expression of any scalar wave can always be written as a linear combination of the elementary modes
\begin{equation}
\Psi(\zeta,\eta,\varphi) =\frac{1}{\sqrt{\bar\zeta\bar\eta}} \Upsilon(\bar\zeta)\Theta(\bar\eta)\Phi(\varphi),\quad\quad
\bar\zeta = \zeta\frac{\omega}{c},\quad\quad\bar\eta = \eta\frac{\omega}{c},
\end{equation}
\begin{equation}
\Phi(\varphi) = \frac{1}{\sqrt{2\pi}} e^{im\varphi},
\end{equation}
\begin{equation}
\frac{d^2\Upsilon}{d\bar\zeta^2} +\Big[\frac{1}{4} + \frac{\kappa}{\bar\zeta}+ \frac{1/4 -(m/2)^2}{\bar\zeta^2}\Big] \Upsilon = 0,\label{eq:upsilon}
\end{equation}

\begin{equation}
\frac{d^2\Theta}{d\bar\eta^2} +\Big[\frac{1}{4} - \frac{\kappa}{\bar\eta}+ \frac{1/4 -(m/2)^2}{\bar\eta^2}\Big] \Theta = 0\label{eq:Theta}.
\end{equation}
The solutions \cite{abra} of the latter equations
 can be written in terms of the Whittaker functions $M_{i\kappa,\mu}$ and $W_{i\kappa,\mu}$ of imaginary argument
\begin{eqnarray}
 z^{-1/2}M_{i\kappa,m/2}(iz) &=&c_{\kappa,m} z^{m/2}e^{-iz/2}M((m +1)/2 - i\kappa, 1+m, iz),\nonumber\\
 z^{-1/2}W_{i\kappa,m/2}(iz)  &=& h_{\kappa,m} z^{m/2}e^{-iz/2}U((m +1)/2 - i\kappa, 1+m, iz).
\end{eqnarray}

For the interior problem, the solution of Eqs.~(\ref{eq:upsilon}) and (\ref{eq:Theta}) that is well behaved is given by  $M_{i\kappa,\mu}$, so that the scalar wave solutions take the form
\begin{equation}
\Psi(\zeta,\eta,\varphi) =\sum_{\kappa,m} a_{\kappa,m} e^{im\varphi}V_{\kappa,\vert m\vert }(\omega\zeta/c)V_{-\kappa,\vert m\vert}(\omega\eta/c),\end{equation}
where the notation
\begin{equation}
V_{\kappa,m}(z) = z^{\vert m \vert/2} e^{-iz/2}M\Big(\frac{\vert m\vert +1}{2} - i\kappa,\vert m\vert +1;iz\Big),
\end{equation}
has been introduced.
Notice that  $\Psi(\zeta,\eta,\varphi)$ will be even  (odd) under the parity transformation  $x_3\rightarrow -x_3$  ---which corresponds to $\zeta\leftrightarrow \eta$---
if $a_{\kappa,m} = a_{-\kappa,m}$ ($a_{\kappa,m} =-a_{-\kappa,m}$).
 
The functions $V_{\kappa,m}$  satisfy the relations (valid in general for $m\ge 1$)
\begin{eqnarray}
\sqrt{z}V_{\kappa,m-1}(z)&=& d_+ V_{\kappa-i/2,m}(z) + d_- V_{\kappa+i/2,m}(z),\label{eq:v1}\\
\sqrt{z}V_{\kappa,m+1}(z)&=&  -i\Big( m+1\Big)\Big[ V_{\kappa+i/2,m}(z) - V_{\kappa-i/2,m}(z)\Big]\label{eq:v2}\\
\partial_z V_{\kappa\pm i/2,m}(z) &=& \Big(\pm\frac{i}{2} \mp\frac{i\kappa}{z}\Big)V_{\kappa \pm i/2,m}(z) +\frac{m}{z}d_\pm V_{\kappa \mp i/2,m}(z)\label{eq:v4}\\
\Big[\frac{2}{m}
\partial_z\sqrt{z} + \frac{1}{\sqrt{z}}\Big]V_{\kappa,m+1}(z) &=& \frac{m+1}{m}\Big[V_{\kappa +i/2,m}(z) + V_{\kappa -i/2,m}(z)\Big]\label{eq:v5}\\
\Big[\frac{2}{m}\partial_z\sqrt{z} - \frac{1}{\sqrt{z}}\Big]V_{\kappa,m-1}(z) &=& \frac{i}{m}\Big[d_-V_{\kappa +i/2,m}(z) -d_+V_{\kappa -i/2,m}(z)\Big]\label{eq:v6}
\end{eqnarray}
with
\begin{equation}\label{eq:dpm}
d_\pm =\frac{1}{2} \pm i\frac{\kappa}{m}.
\end{equation}
Using Kummer transformation,
\begin{equation}
M(a, b, z) = e^z M (b-a,b,-z), 
\end{equation}
it can be directly shown that
\begin{equation}
V_{\kappa-i/2,m}(x) = V_{\kappa+i/2,m}^*(x) \label{eq:real}
\end{equation}
for $\kappa$ and $x$ real variables.

In the case of $\kappa$ a real number and $m$  an odd natural number,
\begin{equation}
V_{\kappa,m}(z) = \frac{
2\Gamma(m+1)e^{\pi\kappa/2}}
{\vert\Gamma(\frac{m+1}{2} + i\kappa)\vert}
 \frac{1}{\sqrt{z}}F_{\frac{m-1}{2}}\Big(\kappa;\frac{z}{2}\Big)\label{eq:coul}\end{equation}
with $F_{\frac{m-1}{2}}$ a real valued Coulomb function. These functions are well studied due to their relevance in the context of the Dirac wave function of an electron in a Coulomb potential \cite{abra}, and as such, provide a guideline for properties ---{\textit{e. g.}}, limiting forms--- that will result useful in forthcoming Sections.

\subsection{Symmetries, angular spectrum and normalization of the scalar modes}
The symmetries behind the separability of the wave equation in parabolic coordinates \cite{boyer} are induced by the generator of
rotations along the $x_3$-axis
\begin{equation}
\widehat{L}_3,\label{eq:j3} 
\end{equation}
and the operator
\begin{equation}
\frac{1}{2}\Big[\{\widehat{L}_1, \widehat{P}_2\} - \{\widehat{L}_2, \widehat{P}_1\}\Big],\label{eq:O2}
\end{equation}
that results from the subtraction of the symmetrized product of the generator of
rotations along the $x_1$-axis and translations along the $x_2$-axis, and the
symmetrized product of the the generator of
rotations along the $x_2$-axis and the translations along the $x_1$-axis;
both generators act on the scalar wave field.
The latter operator is the third component of the operator obtained from the product of the angular and linear momenta operators, $(\widehat{\boldsymbol{L}}\times\widehat{\boldsymbol{P}} - \widehat{\boldsymbol{P}}\times\widehat{\boldsymbol{L}})/2$. It is the kinetic part of the well studied Runge-Lenz vector, which in turn is the generator of a peculiar symmetry associated to  a charged particle in the presence of the Coulomb potential \cite{pauli}; its connection to a parabolic description of such a system is made explicit in Ref.~\cite{itzykson}.

While in the coordinate representation the expression of  operator Eq.~(\ref{eq:O2}) is cumbersome, in the wave vector representation it is quite simple, as can be directly derived from the equations:
\begin{eqnarray}
\frac{c}{\omega}\mathbf{k} &=& \sin\theta_k \cos \varphi_k {\mathbf e}_1 +  \sin\theta_k \sin \varphi_k {\mathbf e}_2 + \cos\theta_k {\mathbf e}_3, \nonumber\\
-i\widehat L_1&\rightarrow& +\sin\varphi_k \partial_{\theta_k}
+\cos\varphi_k \mathrm{ctan}\theta_k \partial_{\varphi_k},\nonumber\\
-i\widehat L_2&\rightarrow& -\cos\varphi_k \partial_{\theta_k}
+\sin\varphi_k \mathrm{ctan}\theta_k \partial_{\varphi_k},\nonumber\\
-i\widehat L_3&\rightarrow&-\partial_{\varphi_k}.
\end{eqnarray}
Notice that the structure of $\mathbf{k}$ is necessary for the fullfilment of Helmholtz equation in the wave vector space.
This yields
\begin{equation}
\frac{1}{2}\Big[\{\widehat{L}_1, \widehat{P}_2\} - \{\widehat{L}_2, \widehat{P}_1\}\Big]\rightarrow i\sin\theta_k\partial_{\theta_k} + i\cos\theta_k, \label{eq:j1p2j2p1}
\end{equation}
when the unit of length is taken $\vert \mathbf{k}\vert^{-1} = c/\omega$ for  waves with frequency $\omega$.
Notice that $i\sin\theta_k\partial_{\theta_k} + i\cos\theta_k = i\partial_{\theta_k}\sin\theta_k$ measures variations on the angle $\theta_k$ of the scalar wave,  taking into account
the scale factor that projects the wave vector to its component perpendicular to the $x_3$-axis.
The normalized eigenstates of both symmetry operators, Eq.~(\ref{eq:j3}) and Eq.~(\ref{eq:j1p2j2p1}),
\begin{eqnarray}
i\widehat{L}_3 f_{\kappa,m} &=& m f_{\kappa,m}\nonumber\\ 
\frac{1}{2}\Big[(\{\widehat L_1, \widehat{P}_2\} - \{\widehat L_2, \widehat{P}_1\})\Big]f_{\kappa,m}&=&
2\kappa f_{\kappa,m}\label{eq:eigen}
\end{eqnarray}
 are given by the expression
\begin{equation}
f_{\kappa,m}(\theta_k,\varphi_k) =\frac{[\tan \theta_k/2]^{-i2\kappa}}{\sin \theta_k} \frac{e^{im\varphi_k}}{2\pi},\label{eq:fs}
\end{equation}
and they satisfy
\begin{equation}
\int_{\mathbb{S}^{(k)}_2} f^*_{\kappa^\prime,m^\prime}(\theta_k,\varphi_k)f_{\kappa,m}(\theta_k,\varphi_k)d\Omega_k= \delta(2\kappa^\prime -2\kappa) \delta_{m,m^\prime}.
\end{equation}
Equations (\ref {eq:eigen}) give a geometric interpretation ---and can induce a dynamical interpretation--- to the separation variables $m$ and $\kappa$. The eigenfunctions $f_{\kappa,m}$ yield  the angular spectra of the internal solutions of Helmoltz equations since,
\begin{eqnarray}
\bar{\psi}_{\kappa,m}(\mathbf{r}) &\equiv& \int_{\mathbb{S}^{(\mathbf{k})}_2} e^{i(\omega/c)(\mathbf{k}\cdot\mathbf{r})}f_{\kappa,m}(\theta_k,\varphi_k) d\Omega_k\label{eq:angspec}\\
&=& a_{\kappa,m} e^{im\varphi}V_{\kappa,\vert m\vert}(\zeta)V_{-\kappa,\vert m \vert}(\eta) \nonumber\\
\end{eqnarray}
with
\begin{equation}a_{\kappa,m} = (i)^{\vert m \vert} \frac{\Gamma(\frac{\vert m\vert+1}{2} + i\kappa)\Gamma(\frac{\vert m\vert +1}{2} - i\kappa)}{\vert\Gamma(\vert m\vert+1)\vert ^2},\label{eq:norm-scalar}
\end{equation} and ${\mathbb{S}^{(\mathbf{k})}_2}$ the  surface of a sphere of radius $c/\omega$ (taken as one) in the $\mathbf{k}$ space.  The evaluation of $a_{\kappa,m}$ takes into account the integral expression for Bessel functions \cite{prudni},  
$$\int_0^\pi e^{pcos x} (\mathrm{tan}(x/2))^{2\nu}J_{2\mu}(c\sin x)dx =$$ \begin{equation}\frac{1}{c}\frac{\Gamma(\nu+\mu+1/2)\Gamma(\mu-\nu+1/2)}{\Gamma(2\mu+1)\Gamma(2\mu+1)}M_{\nu,\mu}(z_+)M_{\nu,\mu}(z_-),\quad\quad z_\pm = p \pm\sqrt{p^2 - c^2}
\end{equation}
in terms of the Whittaker functions, $M_{i\kappa,\mu}(-i\eta) = \pm i e^{\pm \frac{m}{2}\pi i}M_{-i\kappa,\mu}(i\eta)$; here, use was made of the equation \cite {gradshtein}
$$z^{-1/2-\mu} M_{\lambda,\mu}(z) = (-z)^{-1/2-\mu} M_{-\lambda,\mu}(-z).$$ 
%Notice that there is a missprint in the Nagoya reference, which prevents the usage of the analogous equation reported in that paper.

Equations~(\ref{eq:angspec}) and~(\ref{eq:norm-scalar}) guarantee
\begin{equation}
\int_{\mathbb{R}^3}\bar{\psi}_{\kappa^\prime,m^\prime}^*(\mathbf{r})\bar{\psi}_{\kappa,m}(\mathbf{r})d^3\mathbf{r} =\delta(2\kappa - 2\kappa^\prime)\delta_{m,m^\prime}.
\end{equation}

\section{Electric and magnetic fields for systems with parabolic symmetry}

Let us consider the Hertz potential \cite{nisbet}
\begin{eqnarray}
\boldsymbol{\Pi} &=& \boldsymbol{\pi}e^{-i\omega t}\nonumber\\
&=&\Big[\pi_1 \mathbf{e}_1 + \pi_2\mathbf{e}_2 +\pi_3\mathbf{e}_3\Big]e^{-i\omega t},
\end{eqnarray}
with harmonic time dependence and  Cartesian components $\pi_{1,2,3}$ that are interior solutions of the wave equation
$$\nabla^2\pi_i = -(\omega/c)^2 \pi_i,\quad\quad i=1,2,3$$
with frequency $\omega$. The frequency determines the natural unit of time $[\omega^{-1}]$ and the natural unit of length $[c/\omega]$.

The transverse character of the electromagnetic field in the absence of charge sources ($\nabla\cdot \mathbf{E} = 0 =\nabla\cdot \mathbf{B}$) as well as the Faraday law ($\nabla \times \mathbf{E} =i\mathbf{B}$) and the Maxwell displacement equation ($\nabla \times \mathbf{B} =-i\mathbf{E}$) are satisfied if, either

\begin{eqnarray}
\mathbf{E}_{\mathcal{E}} &=&  \mathbf{\nabla}\times\boldsymbol{\pi},\label{eq:Emodea}\\
\mathbf{B}_{\mathcal{E}} &=& -i\mathbf{\nabla}\times\mathbf{E}_{\mathcal{E}},\label{eq:Emodeb}
\end{eqnarray}
or
\begin{eqnarray}
\mathbf{B}_{\mathcal{B}} &=&  \mathbf{\nabla}\times\boldsymbol{\pi}, \label{eq:Bmodea}\\
\mathbf{E}_{\mathcal{B}} &=&  i\mathbf{\nabla}\times\mathbf{B}_{\mathcal{B}} \label{eq:Bmodeb}.
\end{eqnarray}
The modes given by Eqs.~(\ref{eq:Emodea}) and~(\ref{eq:Emodeb}) will be refered as $\mathcal{E}$-modes, and those obtained from Eqs.~(\ref{eq:Bmodea}) and~(\ref{eq:Bmodeb}) as $\mathcal{B}$-modes.

\subsection{Symmetrized elementary electromagnetic modes}

The symmetries exhibited by the wave equation in parabolic coordinates can be used to define
the elementary modes of the electromagnetic field in the presence of boundaries with parabolic geometry. 
The key point to define these modes corresponds to finding vector Hertz potentials that give rise to
electric and magnetic fields that are eigenfunctions of the generators of the transformations associated to the symmetries
mentioned in  Section II.A. For EM fields, these generators take into account the expected vector behavior of $\mathbf{E}$ and $\mathbf{B}$.

Consider an  infinitesimal  rotation by an angle $\delta\varphi$  about any one of the Cartesian  axes.
A vector field with components $\phi_r$ is transformed under an infinitesimal  rotation by an angle $\delta\varphi$  according to the equation \cite{rhorlich} 
\begin{eqnarray}
\phi_r^\prime &=& \phi_r + \delta\varphi\sum_{s=1,2,3}\widehat{M}_{rsij}\phi_s,\label{eq:m1}\\
              &=& \phi_r + \delta\varphi\Big[\widehat{L}_{ij} \phi_r + \sum_{s=1,2,3}\widehat{\mathcal{S}}_{rsij}\phi_s\Big] ,\label{eq:m2}\\
\widehat{L}_{ij}&=&-i(x_i\partial_j - x_j\partial_i),\label{eq:m3}\\
\widehat{\mathcal{S}}_{rsij} &=& -i \Big[\delta_{ri}\delta_{sj} - \delta_{ri}\delta_{sj}\Big],
\end{eqnarray}
where the indices $i,j$ are determined using the Levi Civita tensor $\varepsilon_{tij}$ for  rotation about the $t$-axis.
$\hat{L}_{ij}$ is the orbital angular momentum tensor operator and $\hat{\mathcal{S}}_{rsij}$ the spin-1 tensor operator.

For parabolic geometry ---as mentioned above for a scalar field--- the symmetries are generated by rotations about the $x_3$-axis, and  
by the operator that results from the subtraction of the symmetrized product of the generator of
rotations along the $x_1$-axis and translations along the $x_2$-axis, and the
symmetrized product of the the generator of
rotations along the $x_2$-axis and the translations along the $x_1$-axis.
For vector fields, orbital and intrinsic factors must be incorporated in the transformation of the field, so that,
\begin{eqnarray}
\hat{J}_3 \phi_r &=& -i(x_1\partial_2 - x_1\partial_2)\phi_r + \sum_{s=1,2,3}\widehat{\mathcal{S}}_{rs12}\phi_s,\label{eq:Jz}\\
\hat{\mathfrak{A}}_3\phi_r &=& \frac{1}{2}\Big[\{\widehat{M}_{rs23},\widehat{P}_2\} - \{\widehat{M}_{rs31},\widehat{P}_1\}\Big]\phi_s.
\end{eqnarray}
The symmetrized electromagnetic modes we are looking for  correspond to electric  $\mathbf{E}_{j,\alpha} $ and magnetic $\mathbf{B}_{j,\alpha}$ fields which satisfy Maxwell equations, and are eigenvectors of the operators $\hat{J}_3$ and  $\hat{\mathfrak{A}}_3$,
\begin{eqnarray}
\hat{J}_3 \mathbf{E}_{j,\alpha} &=& j \mathbf{E}_{j,\alpha},\quad\quad \hat{J}_3  \mathbf{B}_{j,\alpha} = j \mathbf{B}_{j,\alpha}\label{eq:evJ}\\
\hat{\mathfrak{A}}_3 \mathbf{E}_{j,\alpha} &=& \alpha\mathbf{E}_{j,\alpha},\quad\quad \hat{\mathfrak{A}}_3 \mathbf{B}_{j,\alpha} = \alpha \mathbf{B}_{j,\alpha} .\label{eq:evA}
\end{eqnarray}
 $\mathbf{E}_{j,\alpha} $ and $\mathbf{B}_{j,\alpha}$ are determined by the Hertz potential according to either Eqs.~(\ref{eq:Emodea}) and~(\ref{eq:Emodeb}) or Eqs.~(\ref{eq:Bmodea}) and~(\ref{eq:Bmodeb}).

Coupling of the orbital and intrinsic angular momenta of the EM field gives rise to electric and magnetic modes that are eigenvectors of $\hat J_3$, Eqs.~(\ref{eq:evJ}).
This coupling can be easily implemented using the circular basis 
\begin{equation}
\mathbf{e}_\pm \equiv \mathbf{e}_1 \pm i \mathbf{e}_2, \quad \quad \mathbf{e}_0 \equiv \mathbf{e}_3,
\end{equation}
to write the Hertz potential
\begin{subequations}\label{eq:summ}
\begin{align}
\boldsymbol{\pi}= \pi_+ \mathbf{e}_+ + \pi_-\mathbf{e}_- +\pi_0\mathbf{e}_0;\\
\pi_{+} = \sum_\kappa c_{\kappa,m}^{(+)}e^{i (m-1)\varphi}V_{\kappa,m-1}(\zeta)V_{-\kappa,m-1}(\eta),\\
\pi_{-} = \sum_\kappa c_{\kappa,m}^{(-)}e^{i (m+1)\varphi}V_{\kappa,m+1}(\zeta)V_{-\kappa,m+1}(\eta),\\
\pi_{0} = \sum_\kappa c_{\kappa,m}^{(0)}e^{i m\varphi}V_{\kappa,m}(\zeta)V_{-\kappa,m}(\eta) .
\end{align}
\end{subequations}
It results that
\begin{equation}
\hat{J}_3 \boldsymbol{\pi}  = m \boldsymbol{\pi},
\end{equation}
with an analogous equation for the derived electric and magnetic fields.

Elementary modes are solutions of the electromagnetic wave equations satisfying the boundary conditions derived from the physical situation under consideration
and with a minimal set of elements in the summations over the labels $\{\omega,m,\kappa\}$ in Eqs.(\ref{eq:summ}).

In the search of the adequate structure of the Hertz potential $\boldsymbol{\pi}$ to fulfill Eqs.~(\ref{eq:evJ}) and~(\ref{eq:evA}) with a finite number of terms  in Eqs.~(\ref{eq:summ}),
 it results convenient to work out its rotational.
Written in terms of the parabolic unit vectors and scale factors
\begin{equation}
\boldsymbol{\pi} = \Big[\frac{P_+}{h_\eta}+\frac{\pi_0}{2h_\zeta}\Big] \mathbf{e}_\zeta  +
\Big[\frac{P_+}{h_\zeta}-\frac{\pi_0}{2h_\eta}\Big] \mathbf{e}_\eta
-iP_-\mathbf{e}_\varphi,\label{eq:vec_m}
\end{equation}
\begin{equation}
P_+=\frac{e^{i\varphi}\pi_+ +e^{-i\varphi}\pi_-}{2},\quad P_-=e^{i\varphi}\pi_+ - e^{-i\varphi} \pi_-.
\end{equation}
From these equations
\begin{eqnarray}
\mathbf{\nabla}\times\boldsymbol{\pi} &=& \frac{p_\zeta}{h_\eta h_\phi} \mathbf{e}_\zeta +
\frac{p_\eta}{h_\zeta h_\varphi} \mathbf{e}_\eta +
\frac{p_\varphi}{h_\eta h_\zeta} \mathbf{e}_\varphi, \nonumber\\
 \mathbf{\nabla}\times(\mathbf{\nabla}\times\boldsymbol{\pi})
 &=& \frac{\mathbf{e}_\zeta}{h_\eta h_\phi}\Big[\frac{\partial}{\partial \eta}\Big[\frac{h_\varphi}{h_\eta h_\zeta}p_\varphi\Big] - i m \frac{h_\eta}{h_\zeta h_\varphi}p_\eta\Big]\nonumber\\ &+&
 \frac{\mathbf{e}_\eta}{h_\zeta h_\phi}\Big[\frac{\partial}{\partial \zeta}\Big[-\frac{h_\varphi}{h_\eta h_\zeta}p_\varphi\Big] + i m \frac{h_\zeta}{h_\eta h_\varphi}p_\zeta\Big]\nonumber\\&+&
 \frac{\mathbf{e}_\varphi}{h_\zeta h_\eta}\Big[\frac{\partial}{\partial \zeta}\Big[-\frac{h_\eta}{h_\zeta h_\varphi}p_\eta\Big] - \frac{\partial}{\partial \eta}\Big[-\frac{h_\zeta}{h_\eta h_\varphi}p_\zeta\Big]\Big],
 \end{eqnarray}
with
 \begin{eqnarray}
p_\zeta &=&i\frac{\partial h_\varphi P_-}{\partial \eta}
-i m\Big[\frac{h_\eta}{h_\zeta}P_+ - \frac{\pi_0}{2}\Big],\nonumber\\
p_\eta &=&-i\frac{\partial h_\varphi P_-}{\partial \zeta}
+i m\Big[\frac{h_\zeta}{h_\eta}P_+ + \frac{\pi_0}{2}\Big],\nonumber\\
p_\varphi&=&\frac{\partial}{\partial\zeta}
\Big[\frac{h_\eta}{h_\zeta}P_+ -\frac{\pi_0}{2}\Big]
- \frac{\partial}{\partial\eta}
\Big[\frac{h_\zeta}{h_\eta}P_+ +\frac{\pi_0}{2}\Big].
\end{eqnarray}

In the expressions for $p_{\eta,\zeta,\varphi}$ we observe the presence of the differential operators
\begin{equation}
\hat{\mathcal{O}}^z_\pm\equiv \partial_z\sqrt{z} \pm \frac{m}{2\sqrt{z}}, \quad\quad z=\eta,\zeta, \label{eq:oz}
\end{equation}
acting on the functions  $V_{\kappa,m}(z)$ contained in Eqs.~(\ref{eq:summ}). Also in these expressions the product of $V_{\kappa,m}(z)$ by $\sqrt{z}$ is frequently found.
As a consequence of Eqs.~(\ref{eq:v1} -\ref{eq:v6}), the functions $V_{\kappa\pm i/2}(z)$ are expected to appear in the expressions of the EM fields obtained from $\boldsymbol{\pi}$; we then make the following compact $ansatz$ for the Hertz potentials 
\begin{subequations}
\begin{align}
&\pi_{+} = c_{\kappa,m}^{(+)}e^{i (m-1)\varphi}V_{\kappa,m-1}(\zeta)V_{-\kappa,m-1}(\eta),\label{eq:pispacea}\\
&\pi_{-} = c_{\kappa,m}^{(-)}e^{i (m+1)\varphi}V_{\kappa,m+1}(\zeta)V_{-\kappa,m+1}(\eta),\\
&\pi_{0} = c_{\kappa+i/2,m}^{(0)}e^{i m\varphi}V_{\kappa+i/2,m}(\zeta)V_{-(\kappa+i/2),m}(\eta)
+ c_{\kappa-i/2,m}^{(0)}e^{i m\varphi}V_{\kappa-i/2,m}(\zeta)V_{-(\kappa-i/2),m}(\eta).\label{eq:pispacec}
\end{align}
\end{subequations}
looking for elementary EM modes that satisfy the eigenvalue Eq.~(\ref{eq:evA}). In the following paragraph the relevance of this $ansatz$ is demonstrated.

The generator of the parabolic symmetry transformation  for scalar fields was found to have a simple expression in wave vector space, Eq.~(\ref{eq:j1p2j2p1}).
Something similar occurs for the vector modes:
in the wave vector representation the generator $\hat{\mathfrak{A}}_3$ of the symmetry transformation for vector fields can be written as
\begin{equation}
\hat{\mathfrak{A}}_3\phi_r = i(\sin\theta_k\partial_{\theta_k} + \cos\theta_k)\phi_r -i\sum_{s=1,2,3}[(\delta_{r2}\delta_{s3} -\delta_{r3}\delta_{s2})k_2 + (\delta_{r1}\delta_{s3} -\delta_{r3}\delta_{s1})k_1]\phi_s.
\end{equation}
In the wave vector space,
\begin{subequations}\label{eq:piwavest}
\begin{align}
&\tilde{\boldsymbol{\pi}} = \tilde{\pi}_{+}\mathbf{e}_+ + \tilde{\pi}_{-}\mathbf{e}_- + \tilde{\pi}_{0}\mathbf{e}_0,\label{eq:piwavesa}\\
&\tilde{\pi}_{+} = \tilde c_{\kappa,m}^{(+)}f_{\kappa,m-1},\\
&\tilde{\pi}_{-}=\tilde c_{\kappa,m}^{(-)}f_{\kappa,m+1},\\
&\tilde{\pi}_{0} = \tilde c_{\kappa+i/2,m}^{(0)}f_{\kappa+i/2,m}+ \tilde c_{\kappa-i/2,m}^{(0)}f_{\kappa-i/2,m},\label{eq:piwavesd}
\end{align}
\end{subequations}
with $f_{\kappa,m}(\theta_k,\varphi_k)$ given by Eq.~(\ref{eq:fs}).
As a consequence,
\begin{eqnarray}\label{eq:base_circ}
\tilde{\mathbf{A}}^{(\mathcal{E})} &=& i\boldsymbol{k}\times\tilde{\boldsymbol{\pi}} = \tilde{A}^{(\mathcal{E})}_{+}\mathbf{e}_+ + \tilde{A}^{(\mathcal{E})}_{-}\mathbf{e}_- + \tilde{A}^{(\mathcal{E})}_{3}\mathbf{e}_3,\nonumber\\
\tilde{A}^{(\mathcal{E})}_{\pm} &=& \tilde{A}^{(\mathcal{E})}_{1}\pm\tilde{A}^{(\mathcal{E})}_{2}=\mp\sin\theta_k e^{\mp i\varphi}\tilde \pi_0 \pm 2\cos\theta_k\tilde\pi_\pm,\nonumber\\
\tilde{A}^{(\mathcal{E})}_{3}&=& -\sin\theta_k[e^{i\varphi}\tilde\pi_+ + e^{-i\varphi}\tilde\pi_-].
\end{eqnarray}
Using trigonometric identities it can be shown that
\begin{equation}
\sin\theta_k \tan\theta_k/2 = 1-\cos\theta_k,\quad\quad \sin\theta_k \tan^{-1}\theta_k/2 = 1+\cos\theta_k,
\end{equation}
and  $\tilde{\mathbf{A}}^{(\mathcal{E})}$ is found to be an analytic function of  $\theta_k$.

For $\tilde{\mathbf{A}}^{(\mathcal{E})}$ to be an eigenvector of $\hat{\mathfrak{A}}_3$ with eigenvalue $\alpha=2\kappa$, the coefficients $\{\tilde c_{\kappa,m}\}$ must satisfy the equation 
\begin{equation}
\tilde c_{\kappa,m}^{(+)} + \tilde c_{\kappa,m}^{(-)} + \tilde c_{\kappa+i/2,m}^{(0)} - \tilde c_{\kappa-i/2,m}^{(0)} =0.\label{eq:acs}
\end{equation}

Demanding that the following vector
\begin{eqnarray}
\tilde{\mathbf{A}}^{(\mathcal{B})} &=& -\boldsymbol{k}\times(\boldsymbol{k}\times\boldsymbol{\tilde{\pi}}) \nonumber\\&=& \tilde{A}^{(\mathcal{B})}_{+}\mathbf{e}_+ + \tilde{A}^{(\mathcal{B})}_{-}\mathbf{e}_- + \tilde{A}^{(\mathcal{B})}_{3}\mathbf{e}_3,
\end{eqnarray}
with
\begin{eqnarray}
\tilde{A}^{(\mathcal{B})}_{\pm} &=& \cos^2\theta_k \tilde\pi_\pm - \sin^2\theta_k e^{\mp i 2\varphi}\tilde \pi_\mp - \sin\theta_k \cos\theta_k\tilde\pi_0,\nonumber\\
\tilde{A}^{(\mathcal{B})}_{3}&=&- \sin\theta_k \cos\theta_k[e^{i\varphi}\tilde\pi_+ + e^{-i\varphi}\tilde\pi_-] + \sin^2\theta_k \tilde\pi_0,
\end{eqnarray}
to be an eigenvector of $\hat{\mathfrak{A}}_3$, leads to the same relationship, Eq.~(\ref{eq:acs}), for the coefficients $\{\tilde c_{\kappa,m}\}$.

Summarizing: the electric and magnetic fields obtained from either expressions Eqs.~(\ref{eq:Emodea}) and~(\ref{eq:Emodeb}) or Eqs.~(\ref{eq:Bmodea}) and~(\ref{eq:Bmodeb}), are eigenvectors of the generator of $\hat J_3$ with eigenvalue $m$ and the generator $\hat{\mathfrak{A}}_3$ with eigenvalue $2\kappa$, whenever the Hertz potentials in wave vector space, Eqs.~({\ref{eq:piwavest}), involve coefficients $\{\tilde c_{\kappa,m}\}$ satisfying Eq.~(\ref{eq:acs}).

\subsection{Scalar product for the electromagnetic modes. Field quantization: photons with parabolic symmetries}

The overlap  of different  EM field modes can be estimated via a scalar product. 
Let us consider a pair of monochromatic EM modes with common frequency $\omega$ and properties labeled by $a$, $b$.
Their scalar product  is defined by
\begin{equation}
\langle {a\vert b}\rangle =\frac{1}{4\pi}\int_{\mathbb{R}^3} d^3x \Big[ \mathbf{E}_{a}^*(\mathbf{x})\cdot\mathbf{E}_{b}(\mathbf{x}) + \mathbf{B}_{a}^*(\mathbf{x})\cdot\mathbf{B}_{b}(\mathbf{x})\Big]. \label{eq:scalrp}
\end{equation}
In the case $a=b$, the integrand corresponds to the time averaged EM energy density,
\begin{equation}
\rho^{EM}_{Energy} =\frac{1}{4\pi}\Big[ \mathbf{E}_{a}^*(\mathbf{x})\cdot\mathbf{E}_{a}(\mathbf{x}) + \mathbf{B}_{a}^*(\mathbf{x})\cdot\mathbf{B}_{a}(\mathbf{x})\Big]. \label{eq:energyden}
\end{equation}
For parabolic modes the involved integrals can be performed more directly when the electric and magnetic fields are expressed in terms of their angular spectrum.
The elementary modes in the presence of parabolic boundaries involve  $\mathbf{E}_{a}$ and $\mathbf{B}_{a}$ with angular spectra derived
from a Hertz potential
\begin{equation}
{\boldsymbol{\pi}} = \int d^3k\delta(\vert\mathbf{k}\vert - \omega/c) e^{i\mathbf{ k}\cdot\mathbf{r}}\big[\tilde \pi_+ \mathbf{e}_+ + \tilde \pi_- \mathbf{e}_- +\tilde \pi_0 \mathbf{e}_0\big]. \nonumber\\
\end{equation} 
with the structure given by Eqs.~({\ref{eq:piwavesa}-\ref{eq:piwavesd}).
Note that -- up to a normalization factor-- two Hertz potentials will lead to the same EM mode if a linear combination of them
can be written as the gradient of a field. In such a case, the two Hertz potentials are related by a gauge transformation. In wave vector space, the particular solution of Eq.~(\ref{eq:acs}) 
\begin{equation}
\tilde c_{\kappa,m}^{(+)} =\tilde c_{\kappa,m}^{(-)} =-\tilde c_{\kappa+i/2,m}^{(0)} =
\tilde c_{\kappa-i/2,m}^{(0)}
\end{equation}
corresponds to
\begin{equation}
\tilde{\boldsymbol{\pi}}_{trivial} = 2{\mathbf{k}}\Big[\frac{e^{im\varphi_k}}{2\pi}\frac{(\tan\theta_k/2)^{-2\kappa i}}{\sin^2\theta_k}\Big] \label{eq:trivial}
\end{equation}
so that ${\mathbf{k}}\times \tilde{\boldsymbol{\pi}}_{trivial}=0$ induces a gauge transformation. The scalar product Eq.~(\ref{eq:scalrp}) is gauge invariant.

A direct calculation shows that the electric fields $\mathbf{E}_{a,b}$ and the magnetic fields  $\mathbf{B}_{a,b}$ contribute equally to $\langle {a}\vert {b}\rangle$ whenever
they are obtained from vectorial Hertz potentials $\boldsymbol{\pi}$ through Eqs.~(\ref{eq:Emodea}-\ref{eq:Emodeb}) or Eqs.~(\ref{eq:Bmodea}-\ref{eq:Bmodeb}). Besides,
\begin{eqnarray}\langle {a}\vert {b}\rangle &=& (2\pi)^2 \delta_{m_a,m_b}\Big[ \delta_1^{(a:b)} \delta (2(\kappa_b - \kappa_a))
+  \delta_2^{(a:b)}\frac{1}{4\sinh (\kappa_b -\kappa_a)\pi} \nonumber\\&+& \delta_3^{(a:b)}\Big[ \delta (2(\kappa_b - \kappa_a)) -\frac{\kappa_b -\kappa_a}{\sinh (\kappa_b -\kappa_a)\pi}
\Big]\Big]\nonumber\end{eqnarray}
with
\begin{eqnarray}
 \delta_1^{(a:b)} &=& (\tilde c^{+*}_{\kappa_a,m} -\tilde c^{-*}_{\kappa_a,m})(\tilde c^{+}_{\kappa_b,m} -\tilde c^{-}_{\kappa_b,m}) +
(\tilde c^{0*}_{\kappa_a+\frac{i}{2},m} +\tilde c^{0*}_{\kappa_a-\frac{i}{2},m})(\tilde c^{0}_{\kappa_b+\frac{i}{2},m} +\tilde c^{0}_{\kappa_b-\frac{i}{2},m}) \, ,\nonumber \\
 \delta_2^{(a:b)} &=& (\tilde c^{+*}_{\kappa_a,m} + \tilde c^{-*}_{\kappa_a,m} +\tilde c^{0*}_{\kappa_a+\frac{i}{2},m}-\tilde c^{0*}_{\kappa_a-\frac{i}{2},m})
 (\tilde c^{+}_{\kappa_b,m} + \tilde c^{-}_{\kappa_b,m} +\tilde c^{0}_{\kappa_b+\frac{i}{2},m}-\tilde c^{0}_{\kappa_b-\frac{i}{2},m})\, , \nonumber \\
 \delta_3^{(a:b)} &=& 2(\tilde c^{0*}_{\kappa_a+\frac{i}{2},m}\tilde c^{0}_{\kappa_b+\frac{i}{2},m}-\tilde c^{0*}_{\kappa_a-\frac{i}{2},m}\tilde c^{0}_{\kappa_b-\frac{i}{2},m})
-(\tilde c^{+*}_{\kappa_a,m} + \tilde c^{-*}_{\kappa_a,m})(c^{0}_{\kappa_b+\frac{i}{2},m}+c^{0}_{\kappa_b-\frac{i}{2},m})\nonumber\\
&-&(\tilde c^{+}_{\kappa_b,m} + \tilde c^{-}_{\kappa_b,m})(c^{0*}_{\kappa_a+\frac{i}{2},m}+c^{0*}_{\kappa_a-\frac{i}{2},m})\, , \nonumber \\
\end{eqnarray}

Notice that Eq.~(\ref{eq:acs}) guarantees that for symmetrized elementary modes:  $\delta_2^{(a:b)} =0 =\delta_3^{(a:b)}$, thus implying the orthogonality between modes with different $m$ or $\kappa$.

By demanding,
 \begin{equation}
 \delta_1^{(a:b)} = \hbar\omega,
\end{equation}
the symmetrized modes can be used to define the electric field operator,
\begin{equation}
\hat{\mathbf{E}}(\mathbf{x},t) = \int d\omega\sum_{a} \Big({\mathbf{E}}_{a}(\mathbf{x}) e^{-i\omega t} \hat{\mathrm{a}}_{a,\omega} +\mathbf{E}^*_{a}(\mathbf{x}) e^{i\omega t} \hat{\mathrm{a}}^\dagger_{a,\omega}\Big)
\end{equation}
and the magnetic field operator,
\begin{equation}
\hat{\mathbf{B}}(\mathbf{x},t) = \int d\omega\sum_{a} \Big({\mathbf{B}}_{a}(\mathbf{x}) e^{-i\omega t} \hat{\mathrm{a}}_{a,\omega} +\mathbf{B}^*_{a}(\mathbf{x}) e^{i\omega t} \hat{\mathrm{a}}^\dagger_{a,\omega}\Big)
\end{equation}
by introducing the creation and annihilation operators,
\begin{equation}
[\hat{\mathrm{a}}_{a,\omega},\hat{\mathrm{a}}^\dagger_{a^\prime,\omega^\prime}] = \delta_{a,a^\prime}\delta(\omega -\omega^\prime).
\end{equation}
According to the results we have shown, the indices summarized with the label  $a$ include the eigenvalues $m$ and $\kappa$ as well as the  procedure by which $\mathbf{E}$ and $\mathbf{B}$
were evaluated from the Hertz potentials, Eqs.~(\ref{eq:Emodea}-\ref{eq:Bmodeb}).

The dynamical variables of the electromagnetic field define its mechanical identity.  These variables can be interpreted as properties of photons 
via the quantization of the EM field in terms of vectorial modes with the adequate space-time dependence. For Cartesian symmetry, the  modes
are properly described by plane waves where the photon frequency $\omega$, wave vector $\mathbf{k}$, and the helicity $\sigma$ ---defined by the projection of the polarization vector on the wave vector--- characterize the photon energy, linear momentum and intrinsic angular momentum.  For EM systems with circular cylindrical symmetry, Bessel modes can be used
to define photons with  frequency $\omega$,  transverse wave vector component $k_\bot$, azymuthal phase quantum number $m$ and  helicity $\sigma$, and relate them to
photon energy, transverse linear momentum, orbital and intrinsic angular momentum respectively \cite{bessel}. In general, via the Noether theorem, the  generator of  a given symmetry of a system
leads to the identification of a dynamical variable that is conserved even under interaction between subsystems. For parabolic geometries it would be important to identify the dynamical operator
directly related to the symmetry generator  $\hat{\mathfrak{A}}_3$. The expected expression can be extrapolated from the results obtained for circular \cite{bessel}, elliptic \cite{mathieu} and parabolic
\cite{weber} cylindrical symmetries.  However, this should be supplemented by the study of the role of the variable $2\hbar\kappa$ in the interaction of EM fields with matter. A deep  theoretical study in
this direction could derive on  experimental proposals similar to those already carried for other geometries \cite{fortagh}.

\section{Vectorial modes in the presence of a parabolic mirror}

An ideal reflecting mirror with parabolic geometry is a paradigmatic optical system. The description of its focusing properties in terms of rays has been known for a long time.
However, the electromagnetic system admits modes with interesting nontrivial configurations both linked to the vector character of the field and the possibility of singularities in its phase.
In this Section, we find the subset of symmetrized vectorial modes that defines the available EM field configurations in the presence of a parabolic mirror.

The magnetic field on the surface of a mirror determined by the condition $\zeta = \zeta_0$ must satisfy the equation
\begin{equation}
\mathbf{B}\cdot\mathbf{e}_\zeta\bigg|_{\zeta = \zeta_0} = 0,
\label{eq:bca}\end{equation}
while the electric field is such that
\begin{eqnarray}
\mathbf{E}\cdot\mathbf{e}_\eta\bigg|_{\zeta = \zeta_0} = 0, \quad \quad 
\mathbf{E}\cdot\mathbf{e}_\varphi\bigg|_{\zeta = \zeta_0} = 0, \label{eq:bcb}
\end{eqnarray}
on such a surface.

For the modes $\{\mathbf{E}_{\mathcal{E}},\mathbf{B}_{\mathcal{E}}\}$ 
the boundary conditions correspond to
\begin{equation}
 p_\eta \bigg|_{\zeta = \zeta_0}= 0, \quad p_\varphi\bigg|_{\zeta = \zeta_0} = 0,  \label{eq:E_E1}
\end{equation}
for $\zeta = \zeta_0$ and any $\eta$ and $\varphi$, which imply directly
\begin{equation}
 \mathbf{B}\cdot \mathbf{e}_\zeta\bigg|_{\zeta = \zeta_0} = 0 \label{eq:E_E2}
\end{equation}
on the mirror surface.

For the modes $\{\mathbf{E}_{\mathcal{B}},\mathbf{B}_{\mathcal{B}}\}$ 
the boundary conditions become
\begin{eqnarray}
p_\zeta\bigg|_{\zeta = \zeta_0} &=& 0,\label{eq:bcneI}\\
 \frac{\partial}{\partial \zeta}\Big[-\frac{h_\varphi}{h_\eta h_\zeta}p_\varphi\Big] + i m \frac{h_\zeta}{h_\eta h_\varphi}p_\zeta\bigg|_{\zeta = \zeta_0}&=&0,\label{eq:bcneII}\\
 \frac{\partial}{\partial \zeta}\Big[-\frac{h_\eta}{h_\zeta h_\varphi}p_\eta\Big] - \frac{\partial}{\partial \eta}\Big[-\frac{h_\zeta}{h_\eta h_\varphi}p_\zeta\Big] \bigg|_{\zeta = \zeta_0}&=&0,\label{eq:bcneIII}
\end{eqnarray}
on the mirror surface. While at first sight Eqs.~(\ref{eq:bcneI} -
\ref{eq:bcneIII}) seem to be independent, they are not; this can be seen from the following general argument. Since $$\nabla\times(\nabla\times\boldsymbol{\pi}) = \nabla(\nabla\cdot\boldsymbol{\pi}) -\nabla^2\boldsymbol{\pi} = \nabla(\nabla\cdot\boldsymbol{\pi})+ (\omega/c)^2\boldsymbol{\pi}, $$
demanding
\begin{equation}\mathbf{e}_\zeta\times(\nabla\times(\nabla\times\boldsymbol{\pi}))\bigg|_{\zeta = \zeta_0}=0 \label{eq:botazeta}\end{equation} 
for $\zeta = \zeta_0$ and any value of $\eta$ and $\varphi$, guarantees that the tangential derivatives of these quantities are also zero at $\zeta=\zeta_0$, and as a consequence
$$ \mathbf{e}_\zeta\cdot\nabla\times\boldsymbol{\pi}\bigg|_{\zeta = \zeta_0} =
\mathbf{e}_\zeta\cdot\nabla\times(\nabla\times(\nabla\times\boldsymbol{\pi})-\nabla(\nabla\cdot\boldsymbol{\pi} ))\bigg|_{\zeta = \zeta_0}=0.$$

Equation (\ref{eq:botazeta}) can be expressed in terms of $P_{\pm}$ and $\pi_0$. The resulting pair of equations seems to be difficult to solve until it is noticed that: (i) they must be satisfied simultaneously  for $\zeta = \zeta_0$ and any value of $\varphi$ and $\eta$, and  (ii) $\sqrt{\eta}V_{-\kappa,m}(\eta)$ satisfies Whittaker equation. These facts lead to the simplified equations
\begin{eqnarray}
\Big[\frac{\partial\sqrt{\zeta\eta}P_-}{\partial \eta} - \frac{m}{\eta}\sqrt{\zeta\eta}P_++m\frac{\pi_0}{2}\Big]\bigg|_{\zeta =\zeta_0} & = & 0, \label{eq:E_B1}\\
\Big[\frac{\partial\sqrt{\zeta\eta}P_+}{\partial\zeta} +\zeta\frac{\partial \pi_0/2}{\partial\zeta}
-\Big[\frac{\kappa}{m} + \frac{m}{4\zeta} -\frac{\zeta}{4m}\Big]\sqrt{\zeta\eta}P_-\Big]\bigg|_{\zeta=\zeta_0} &=&0. \label{eq:E_B2}
\end{eqnarray}

The general form of any electromagnetic wave in the presence of the parabolic mirror can be written as a linear combination of both the modes  $\{\mathbf{E}_{\mathcal{E}},\mathbf{B}_{\mathcal{E}}\}$ and $\{\mathbf{E}_{\mathcal{B}},\mathbf{B}_{\mathcal{B}}\}$ that satisfy the conditions given by Eqs.~(\ref{eq:bca}) and~(\ref{eq:bcb}).

\subsection{Symmetrized modes in the presence of mirrors.}

The boundary conditions imposed by an ideal parabolic mirror can be studied using the Hertz potentials given by Eq.~(\ref{eq:summ}) as summations over terms containing the functions $V_{\kappa,m}$.
This gives rise to a hierarchy of equations described in detail in Appendix B for the $\mathcal{E}$-modes. Two different scenarios are identified. The first corresponds to modes with a null eigenvalue of $\hat J_3$. In such a case, the indices of the functions $V_{\kappa,m}$ in the Hertz potentials are just $m = 0,1$. The second scenario  corresponds to a non-zero eigenvalue of $\hat{J}_3$. Then, the relevant $m$-values in the $V_{\kappa,m}$ are three, $\vert m-1\vert$, $\vert m+1\vert$ and $\vert m\vert$.
The complexity of the equations for the coefficients of the $V_{\kappa,m}$ functions illustrates the consequences of working with modes that are only partially symmetrized when ignoring the relevance of $\hat{\mathfrak{A}}_3$. 

In the following paragraphs we derive the closed expressions of the {\it fully} symmetrized elementary modes in the presence of an ideal mirrors.

\subsubsection{Parabolic Neuman $m=0$ EM modes}

For $m=0$ we show here that the $\mathcal{E}$-modes can be derived from a simple symmetrized vectorial Hertz potential 
\begin{equation}\boldsymbol{\pi}_{m=0} =
 c_{0}[e^{-i\varphi}\mathbf{e}_+ -e^{i\varphi} \mathbf{e}_-]V_{\kappa_0,1}(\zeta)V_{-\kappa_0,1}(\eta).\label{eq:pi0}\end{equation}
This potential yields an electric field
$$\mathbf{E}_N =\nabla \times \boldsymbol{\pi}_{m=0}$$
which is an eigenvector of $\hat J_3$ and $\hat{\mathfrak{A}}_3$. 
The explicit expression for $\mathbf{E}_N$ is
\begin{equation}
\mathbf{E}_N = 2c_0\frac{\mathbf{e}_\zeta}{h_\eta h_\varphi}\Big[\sqrt{\zeta}V_{\kappa_0,1}(\zeta) \frac{\partial}{\partial\eta}\sqrt{\eta}V_{-\kappa_0,1}(\eta)\Big]
-2c_0\frac{\mathbf{e}_\eta}{h_\zeta h_\varphi}\Big[\sqrt{\eta}V_{-\kappa_0,1}(\eta) \frac{\partial}{\partial \zeta}\sqrt{\zeta}V_{\kappa_0,1}(\zeta)\Big].
\end{equation}
while the accompanying magnetic field $\mathbf{B}_N$ is
\begin{equation}
\mathbf{B}_N = 2c_0  V_{\kappa_0,1}(\zeta)V_{-\kappa_0,1}(\eta) \mathbf{e}_\varphi
\end{equation}
Since $\mathbf{e}_\zeta$ and $\mathbf{e}_\eta$ are superpositions of the radial $\mathbf{e}_\rho=\cos\varphi \mathbf{e}_1 +\sin\varphi \mathbf{e}_2$ and $\mathbf{e}_3$
vectors, Eq.~(\ref{eq:paravec}) in Appendix A, the boundary condition
$$\mathbf{E}_{N}\cdot\mathbf{e}_\varphi = 0,$$
is directly satisfied.
Meawhile, the boundary condition
$$\mathbf{E}_{N}\cdot\mathbf{e}_\eta\bigg|_{\zeta =\zeta_0} = 0$$
is equivalent to the Neuman like equation \begin{equation} \frac{\partial}{\partial\zeta}\sqrt{\zeta}V_{\kappa_0,1} = 0. \label{eq:neu}\end{equation} 
It can be shown that this equation can also be written as
\begin{equation}\label{eq:neu}
\mathcal{W}_{\kappa,0}(\zeta_0) \equiv\frac{V_{\kappa+i/2,0}(\zeta_0)}{V_{\kappa-i/2,0}(\zeta_0)} = -1.
\end{equation}

Parabolic Neuman modes were already studied in Refs.~\cite{noeckel}-\cite{luis}. The $\kappa=0$ mode yields an electric field that resembles
that produced by an electric dipole.  

In Figure \ref{fig:neu} we illustrate the properties of Neuman modes with small values of $\kappa$. The EM field is highly focused. Notice that
the contribution of the electric and magnetic field to the EM energy density, Eq.~(\ref{eq:energyden}), is not balanced between those fields; that is, contrary to a plane wave, $\vert \mathbf{E}(\mathbf{r},t)\vert$
may be different to $\vert \mathbf{B}(\mathbf{r},t)\vert$. An optical vortex is located at the origin for the radial component of the electric field, and a high similarity to a radial doughnut mode \cite{doughnut} near the focus of the mirror can also be observed.

\begin{figure}
\begin{tabular}{@{}c @{}c @{}c}
\includegraphics[width=0.335\textwidth,trim= 2cm 6cm 0cm 2cm,clip=true]{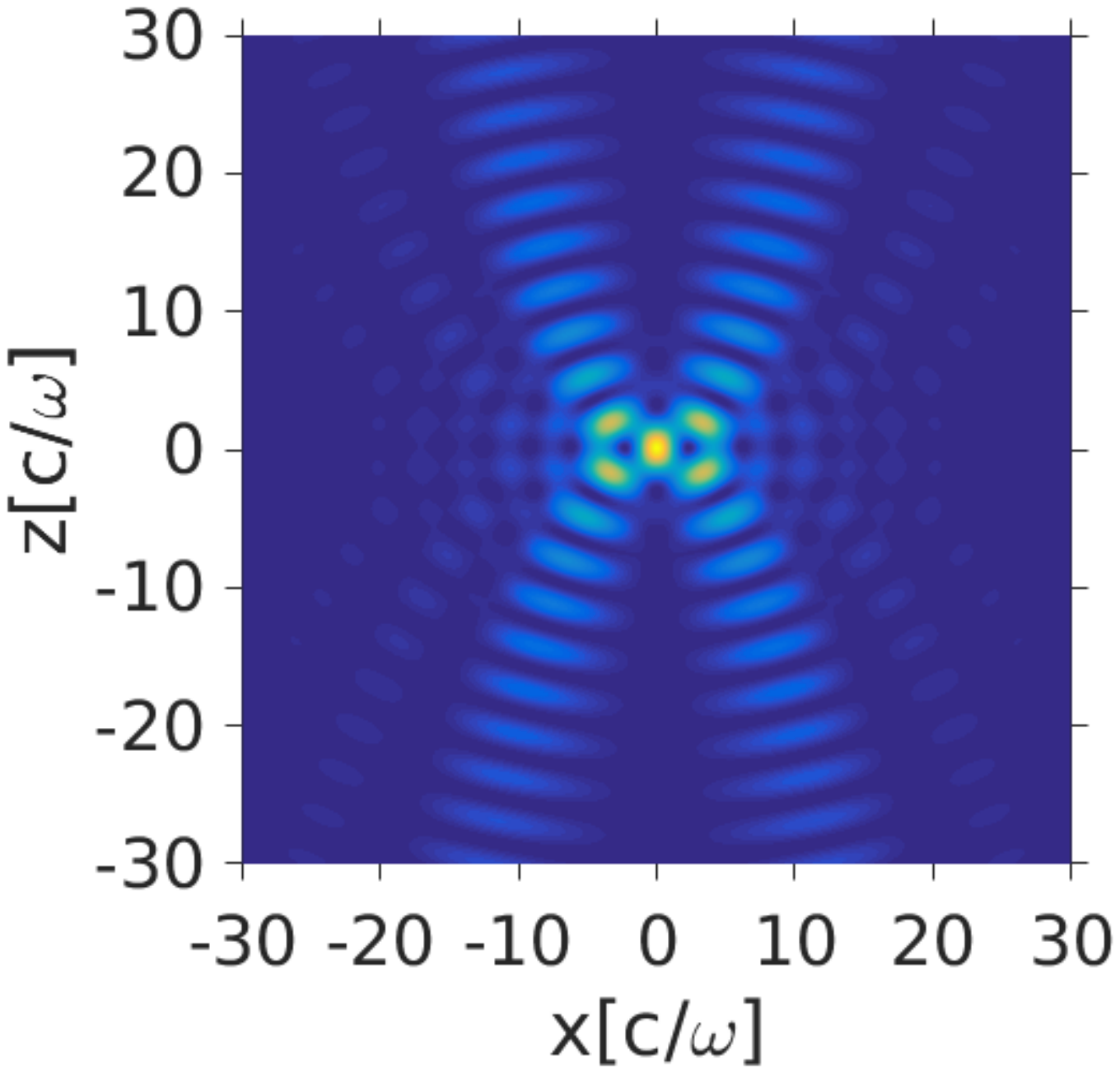}&
 \includegraphics[width=.335\textwidth,trim= 2cm 6cm 0cm 2cm,clip=true]{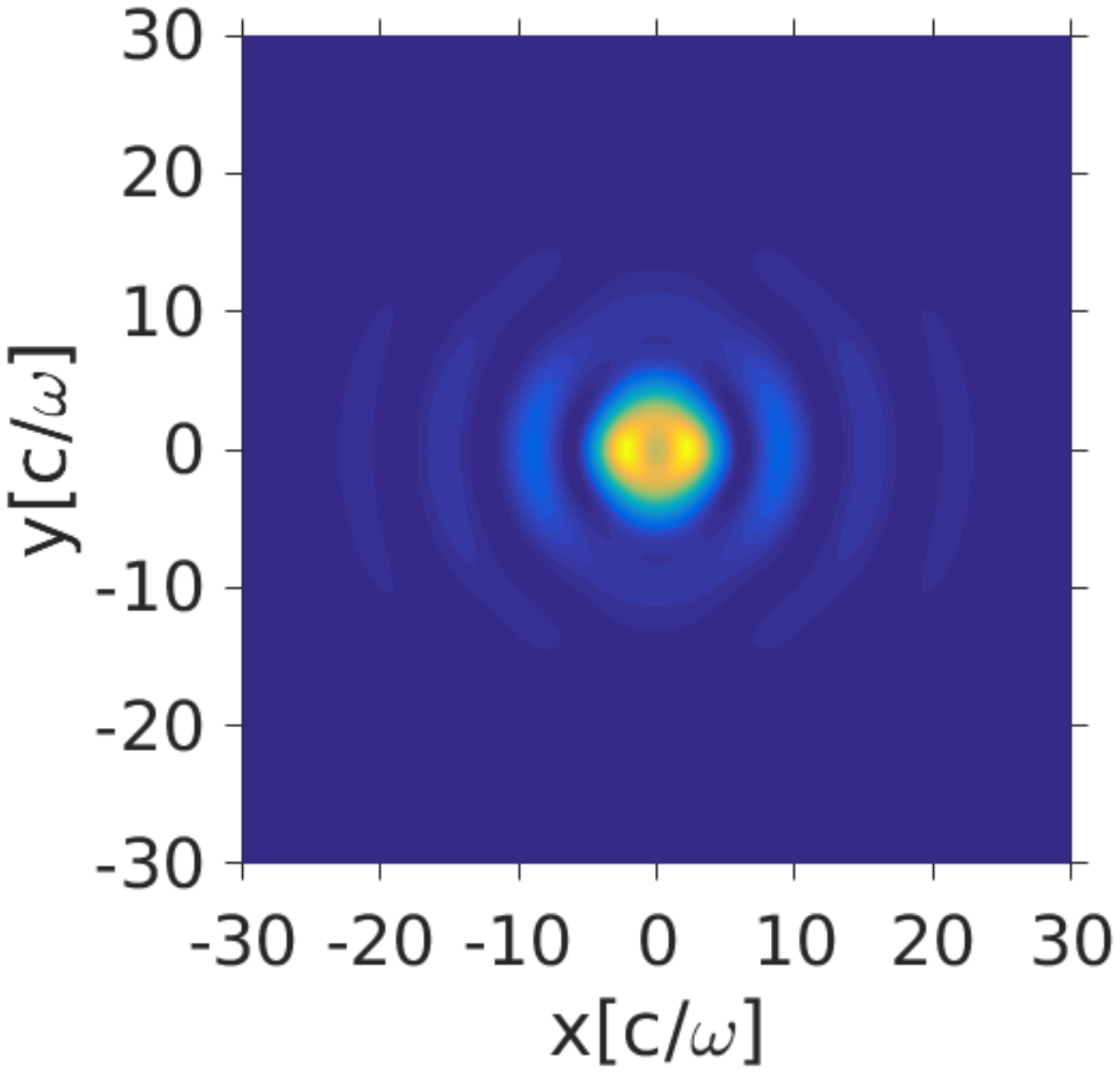}&
\includegraphics[width=.335\textwidth,trim= 2cm 6cm 0cm 2cm,clip=true]{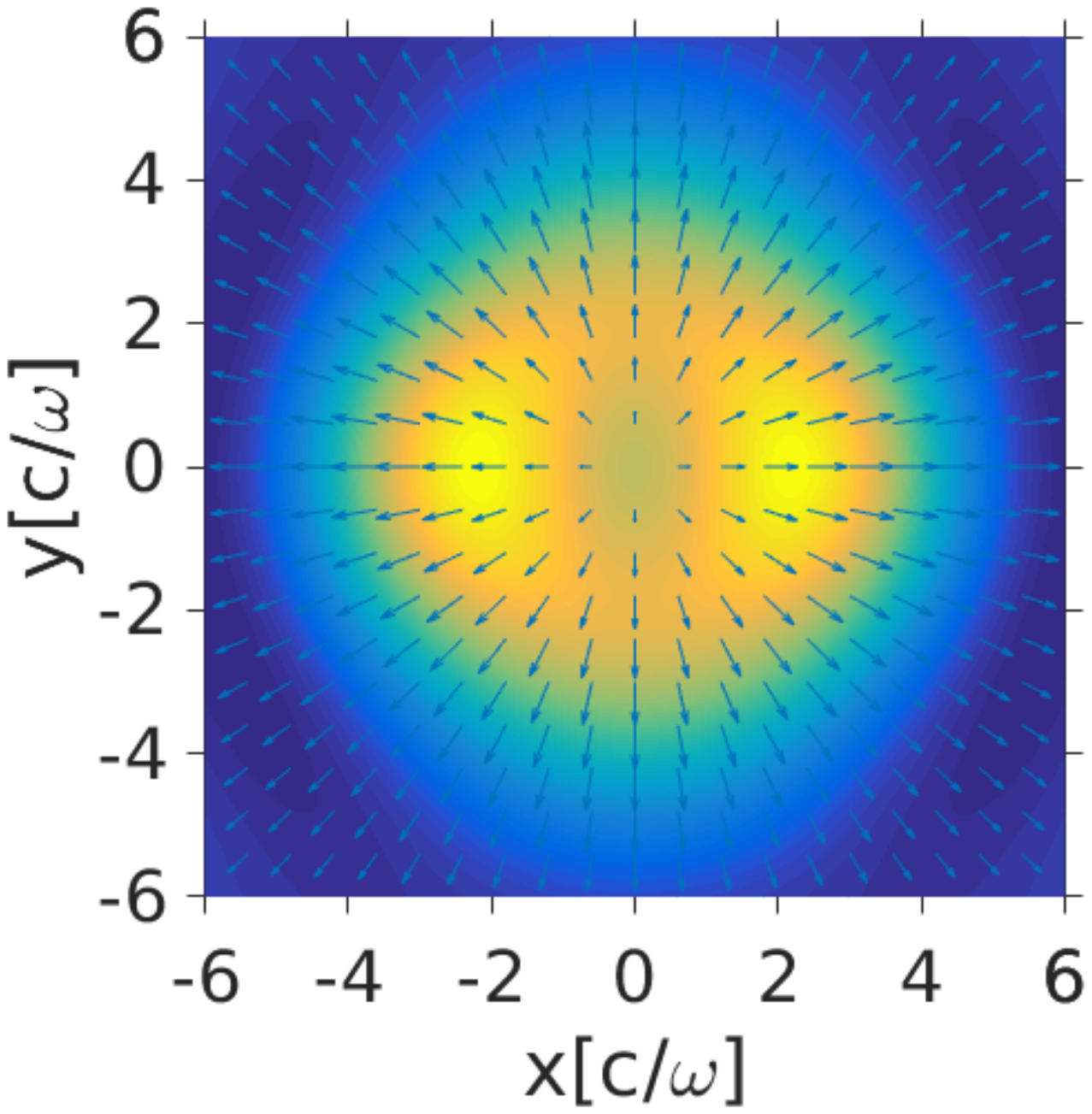}\\
(a) &(b) & (c) \\
\includegraphics[width=0.335\textwidth,trim= 2cm 6cm 0cm 2cm,clip=true]{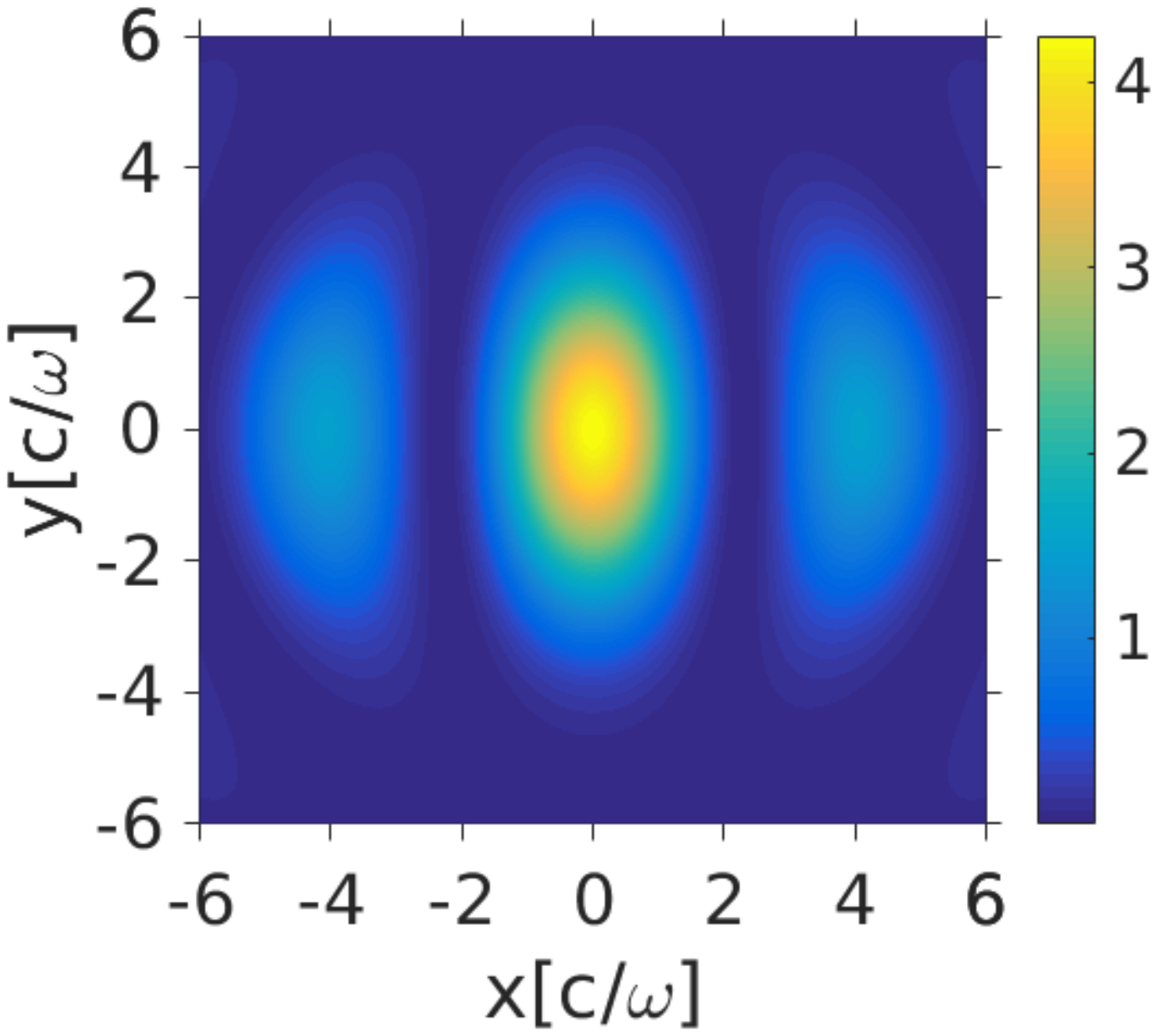}&
\includegraphics[width=0.335\textwidth,trim= 2cm 6cm 0cm 2cm,clip=true]{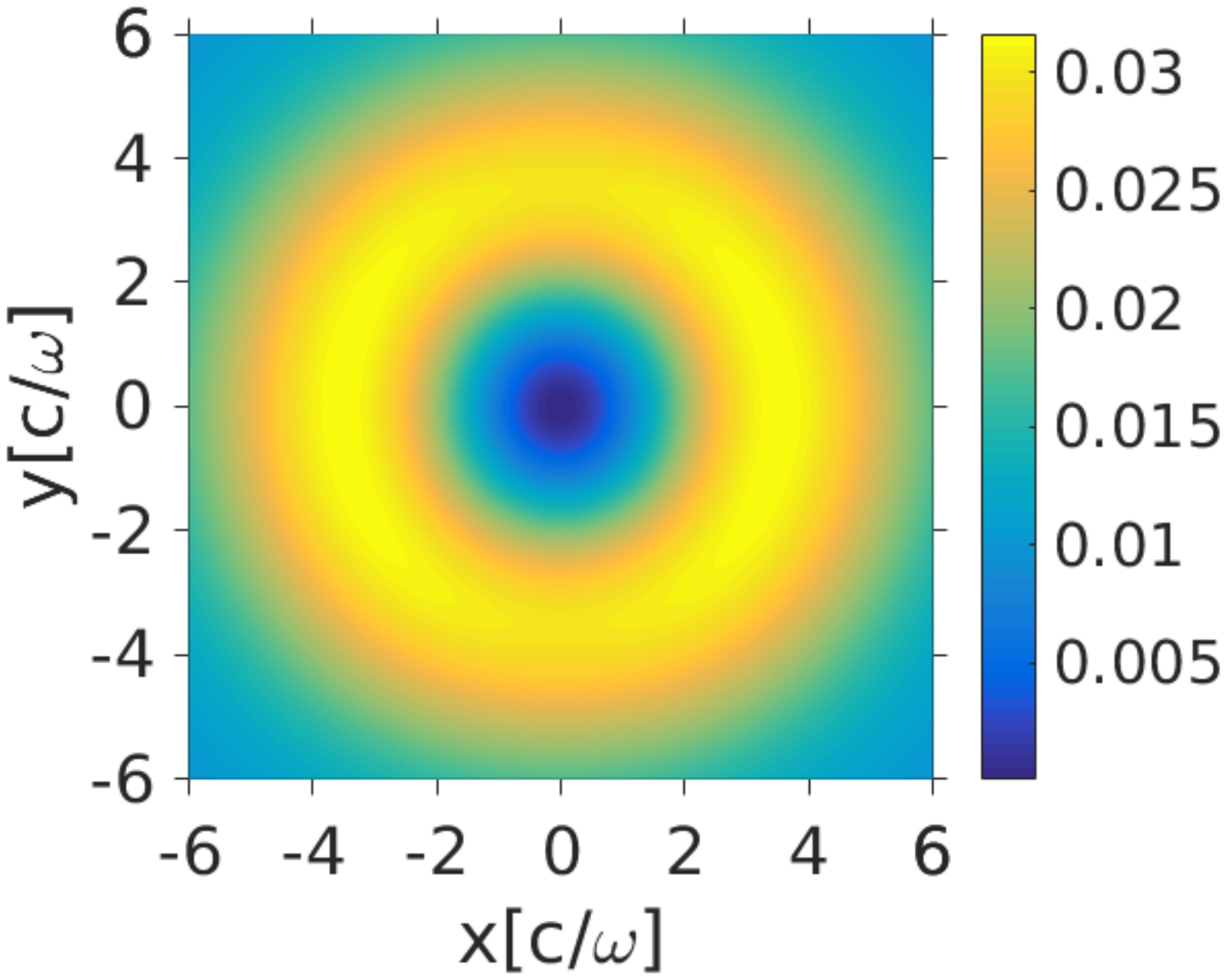}&
\includegraphics[width=.335\textwidth,trim= 2cm 6cm 0cm 2cm,clip=true]{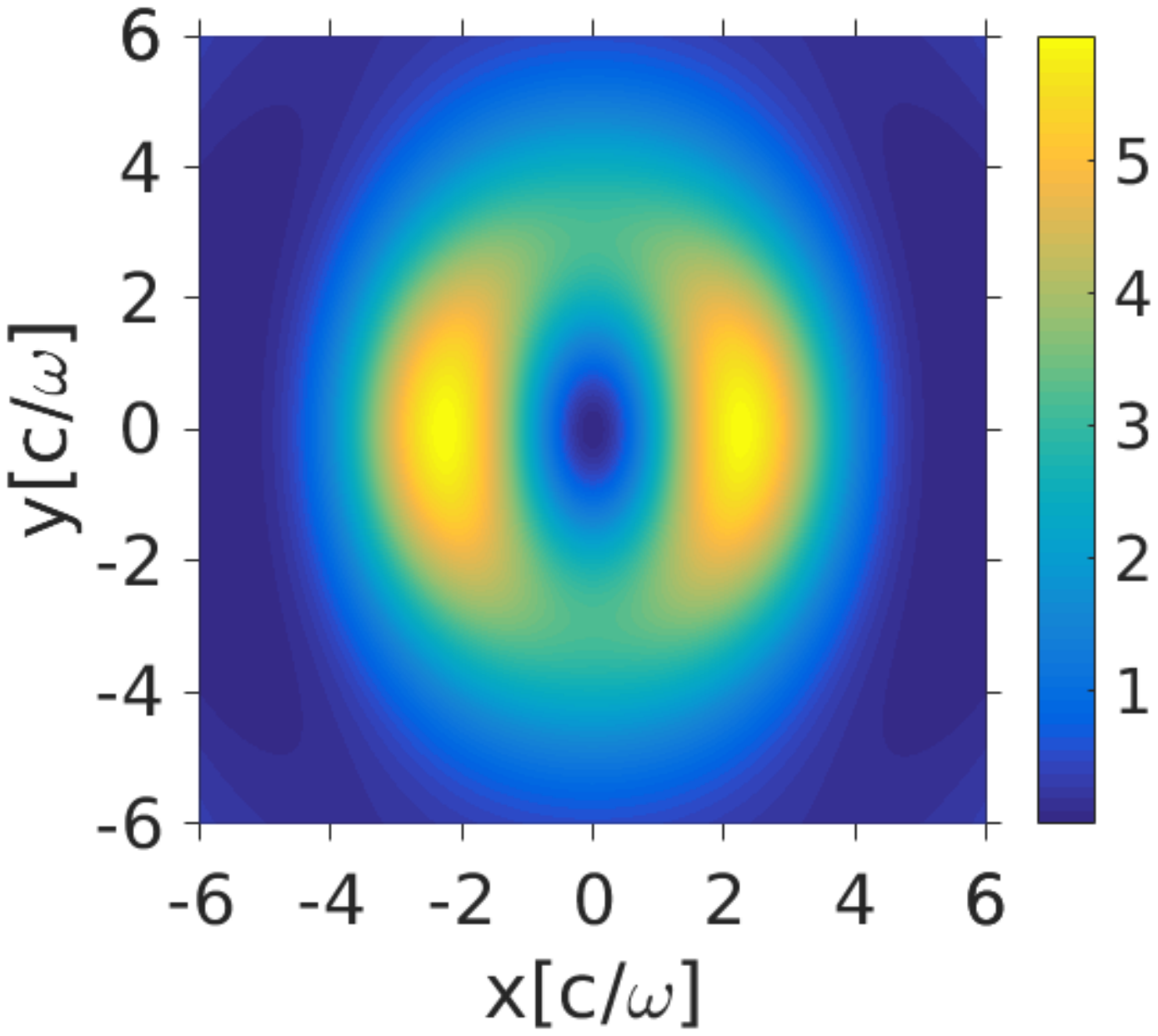}\\
 (d) & (e) & (f)\\
\includegraphics[width=0.335\textwidth,trim= 2cm 6cm 0cm 2cm,clip=true]{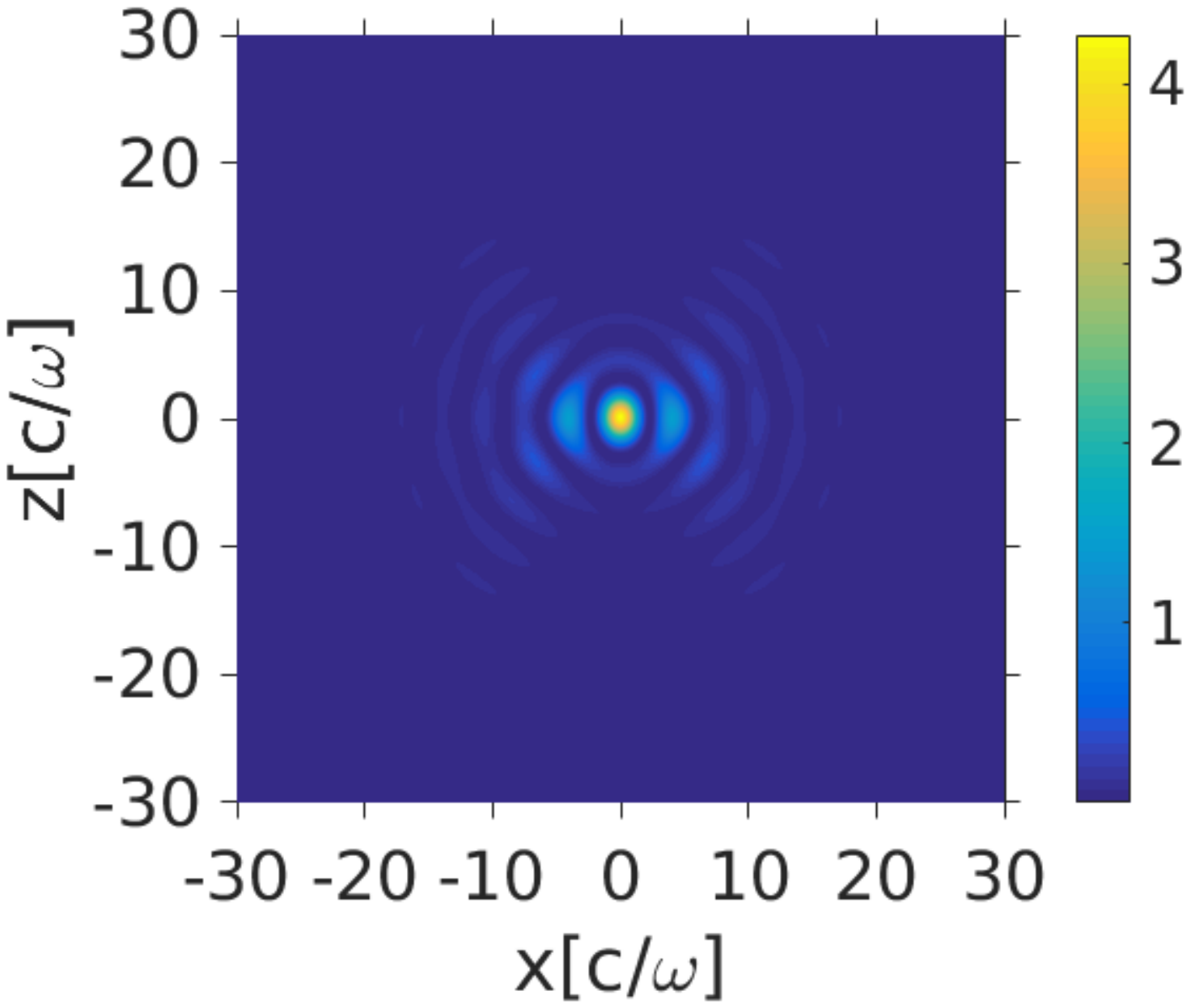}&
\includegraphics[width=0.335\textwidth,trim= 2cm 6cm 0cm 2cm,clip=true]{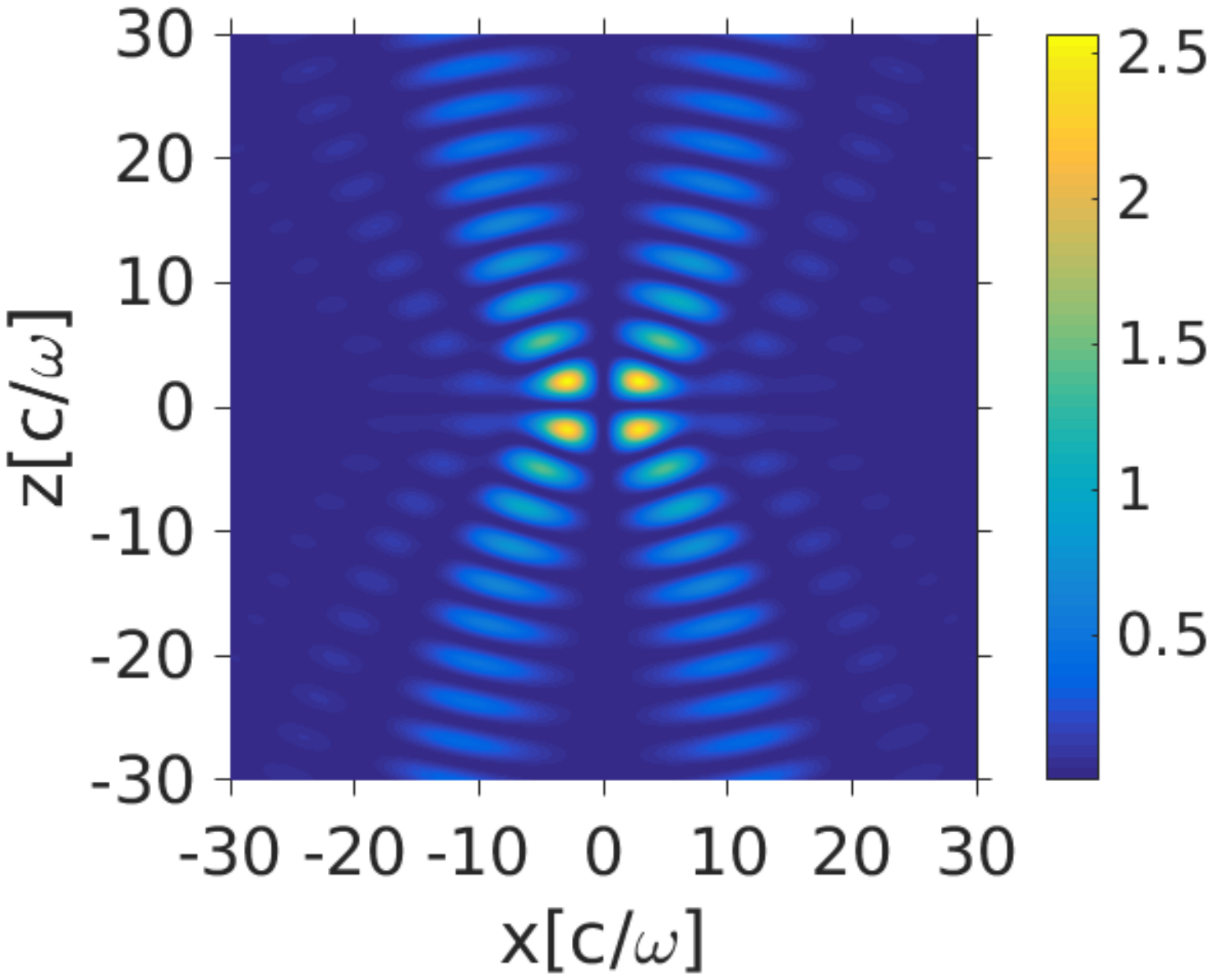}&
\includegraphics[width=.335\textwidth,trim= 2cm 6cm 0cm 2cm,clip=true]{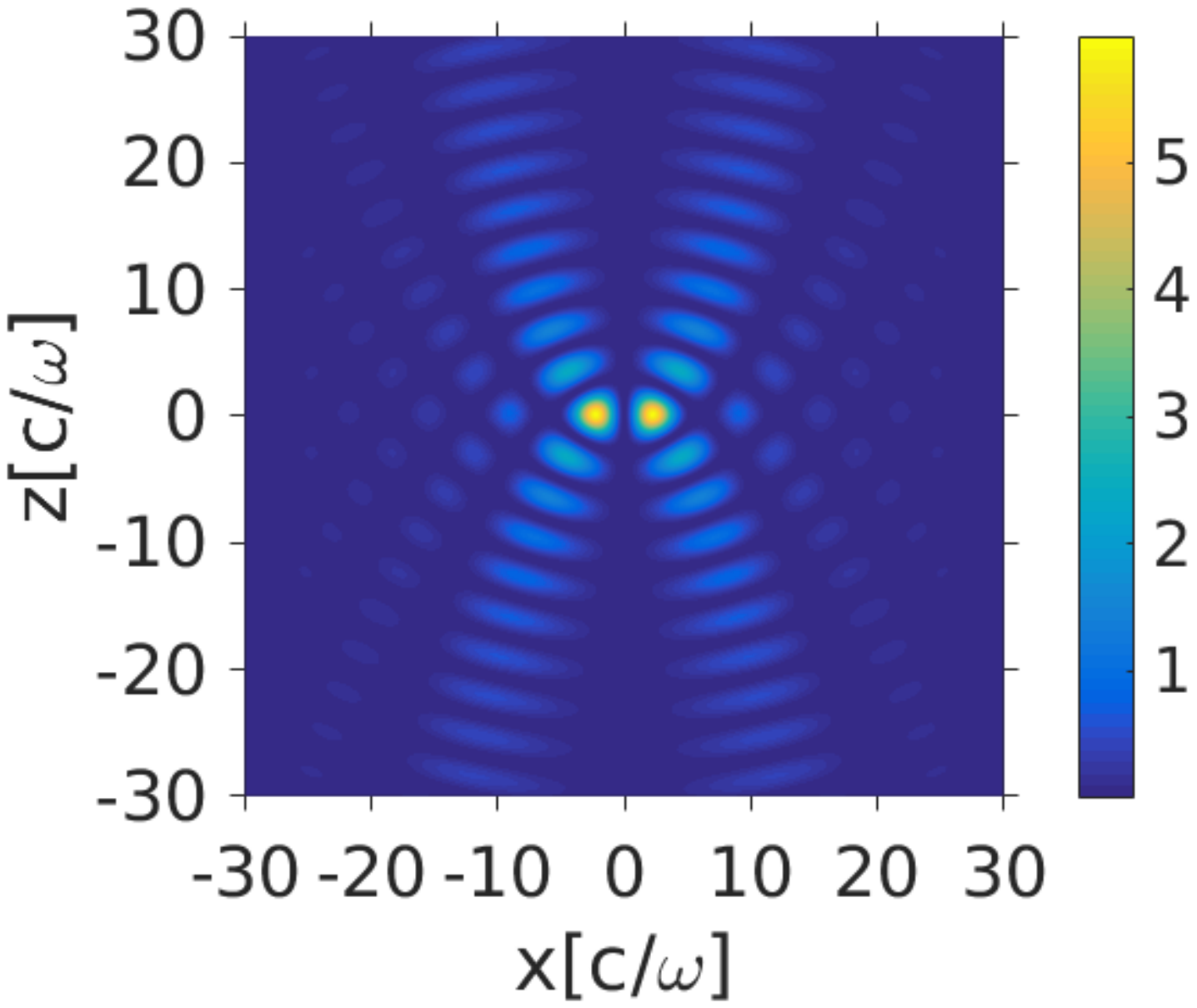}\\
 (d) & (e) & (f)\\
\end{tabular} 
\caption{ Illustrative example of the density of EM energy Eq.~(\ref{eq:energyden}), at the (a) plane defined by $x_1=0$ and at the (b) plane defined by $x_3=0$; (c) illustrates the projection of the real part of the electric field on  the  plane defined by $x_3=0$;
in (d) $\vert \mathbf{E}\cdot\mathbf{e}_3\vert^2 $, (e)  $\vert\mathbf{E}\times\mathbf{e}_3\vert^2 $ and (f) $\vert \mathbf{B}\vert ^2$ are shown at such  plane;
in (g) $\vert \mathbf{E}\cdot\mathbf{e}_3\vert^2 $, (h)  $\vert\mathbf{E}\times\mathbf{e}_3\vert^2 $ and (i) $\vert \mathbf{B}\vert ^2$ are shown at the $x_2 = 0$ plane.
 All subfigures correspond to a Neuman mode with the parameter $\kappa= -.040937$ and the coefficient $\tilde c_0 =1$. This  $\kappa$ is the root with smallest absolute value of the Neuman boundary condition, Eq.~(\ref{eq:neu}), for a mirror surface at $\zeta_0 = 35700 c/\omega$. }\label{fig:neu}
\end{figure}

\subsubsection{Parabolic Dirichlet $m=0$ EM modes}

Consider now a $\mathcal{B}$-mode, derived from a Hertz potential with the structure Eq.~(\ref{eq:pi0}),
$$\mathbf{E}_D =\nabla\times(\nabla\times\boldsymbol{\pi}_{m=0})$$
Then,
\begin{equation}
\mathbf{E}_D = E_0 V_{\kappa_0,1}(\zeta)V_{-\kappa_0,1}(\eta)\mathbf{e}_\varphi\nonumber\\
\end{equation}
so that,
$$\mathbf{E}_D\cdot\mathbf{e}_\eta = 0.$$
The Dirichlet condition 
\begin{equation}
V_{\kappa_0,1}(\zeta_0)= 0 \label{eq:diri}
\end{equation}
is necessary for satisfying
$$\mathbf{E}_D\cdot\mathbf{e}_\varphi\bigg|_{\zeta =\zeta_0} = 0.$$
Equation~(\ref{eq:diri}) is equivalent to
\begin{equation}\label{eq:dirichlet}
\mathcal{W}_{\kappa,0}(\zeta_0) \equiv\frac{V_{\kappa+i/2,0}(\zeta_0)}{V_{\kappa-i/2,0}(\zeta_0)} = 1.
\end{equation}

The polarization of the Dirichlet modes is orthogonal to that of the Neuman modes.

In Figure \ref{fig:dir} the Dirichlet modes with small value of $\vert \kappa\vert $ are illustrated. We have choosen the same location of the parabolic mirror than
the one used in Figure \ref{fig:neu}. Though the values of $\kappa$ for the illustrated Neuman and Dirichlet modes are similar, their general structure and
configurations for their $\mathbf{E}$ and $\mathbf{B}$ fields, are very different near the focus.
\begin{figure}
\begin{tabular}{@{}c @{}c @{}c}
\includegraphics[width=0.335\textwidth,trim= 2cm 6cm 0cm 2cm,clip=true]{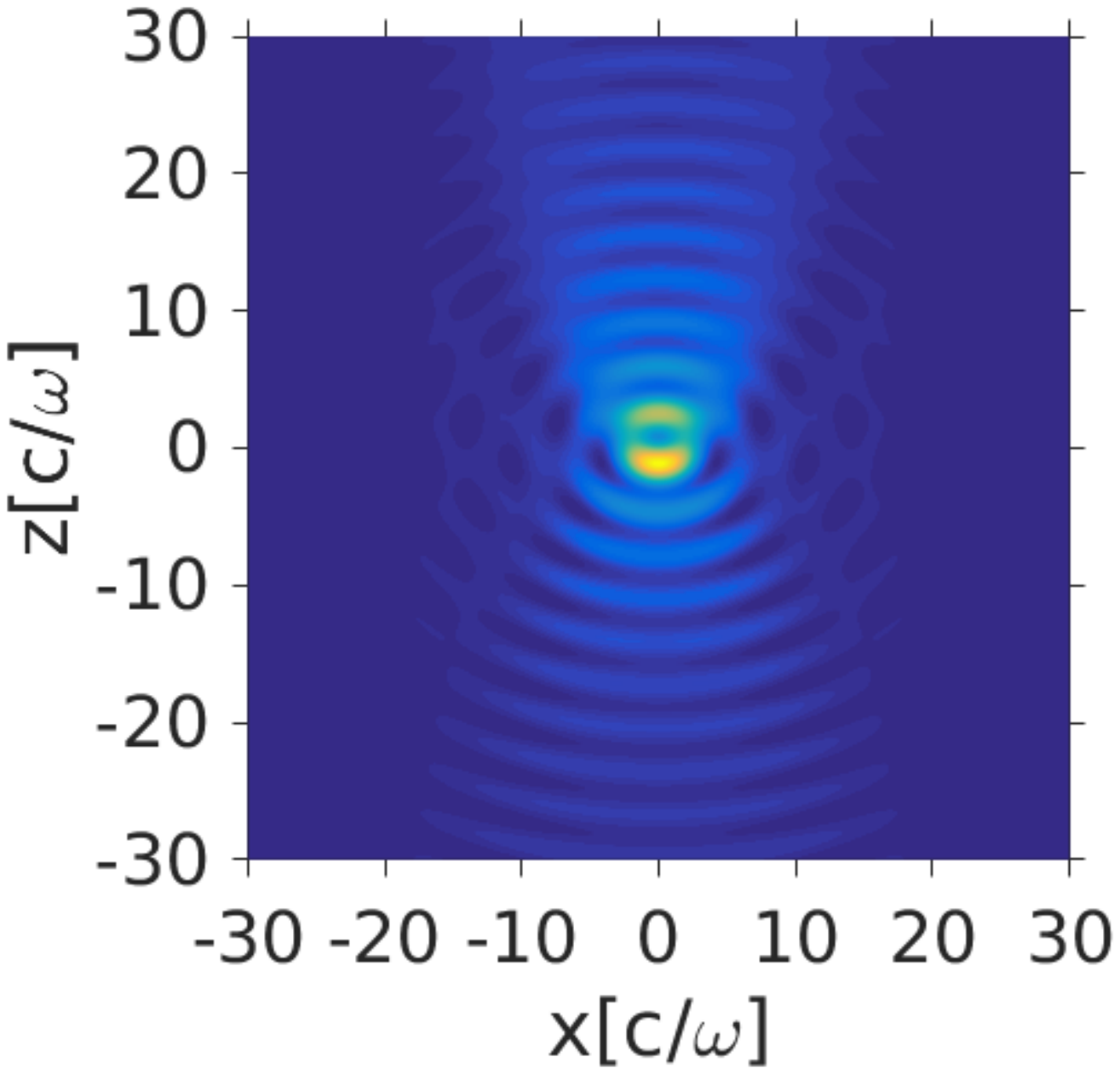}&
\includegraphics[width=0.335\textwidth,trim= 2cm 6cm 0cm 2cm,clip=true]{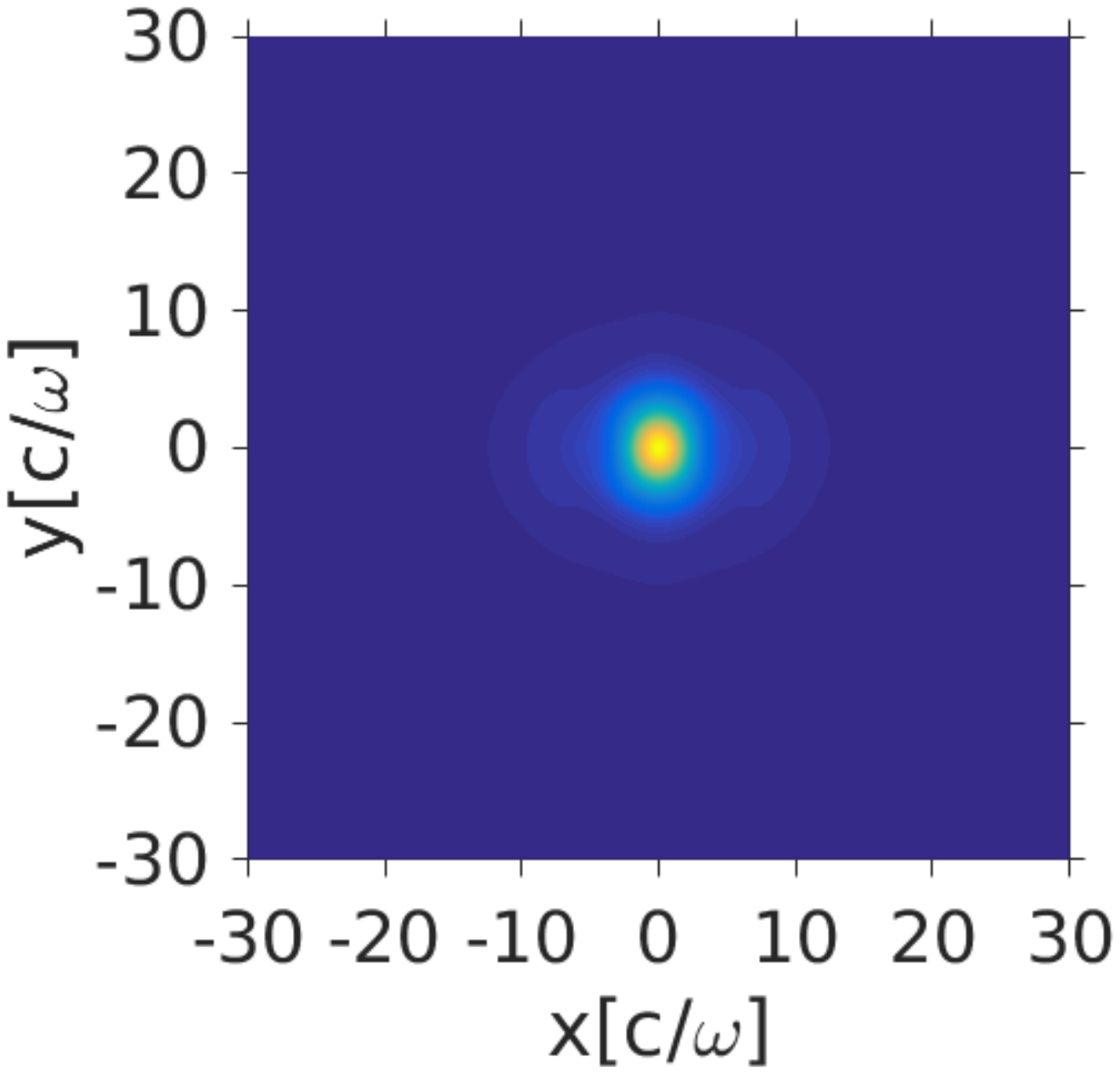}&
 \includegraphics[width=.335\textwidth,trim= 2cm 6cm 0cm 2cm,clip=true]{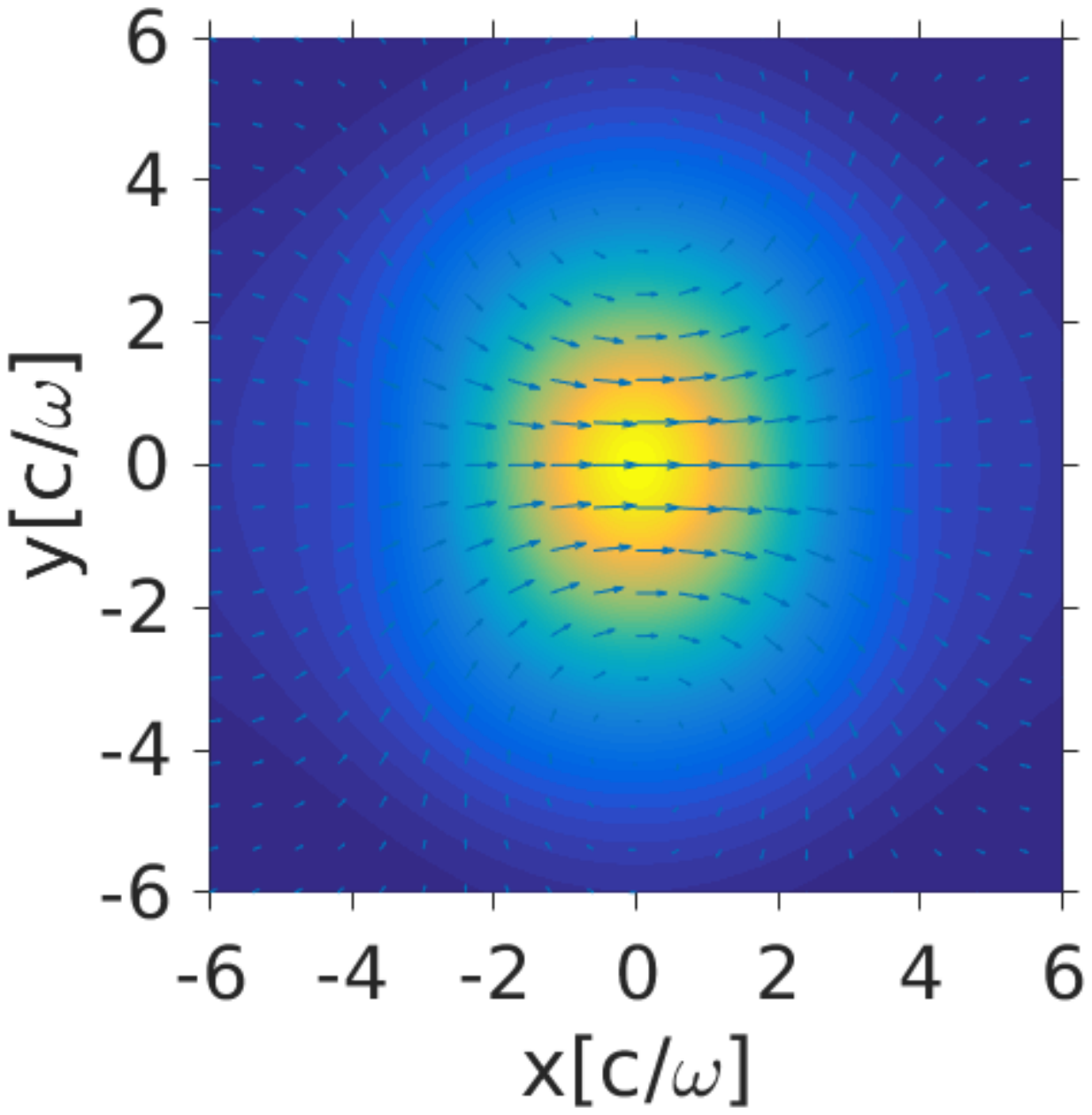}\\
(a) &(b) &(c) \\
\includegraphics[width=0.335\textwidth,trim= 2cm 6cm 0cm 2cm,clip=true]{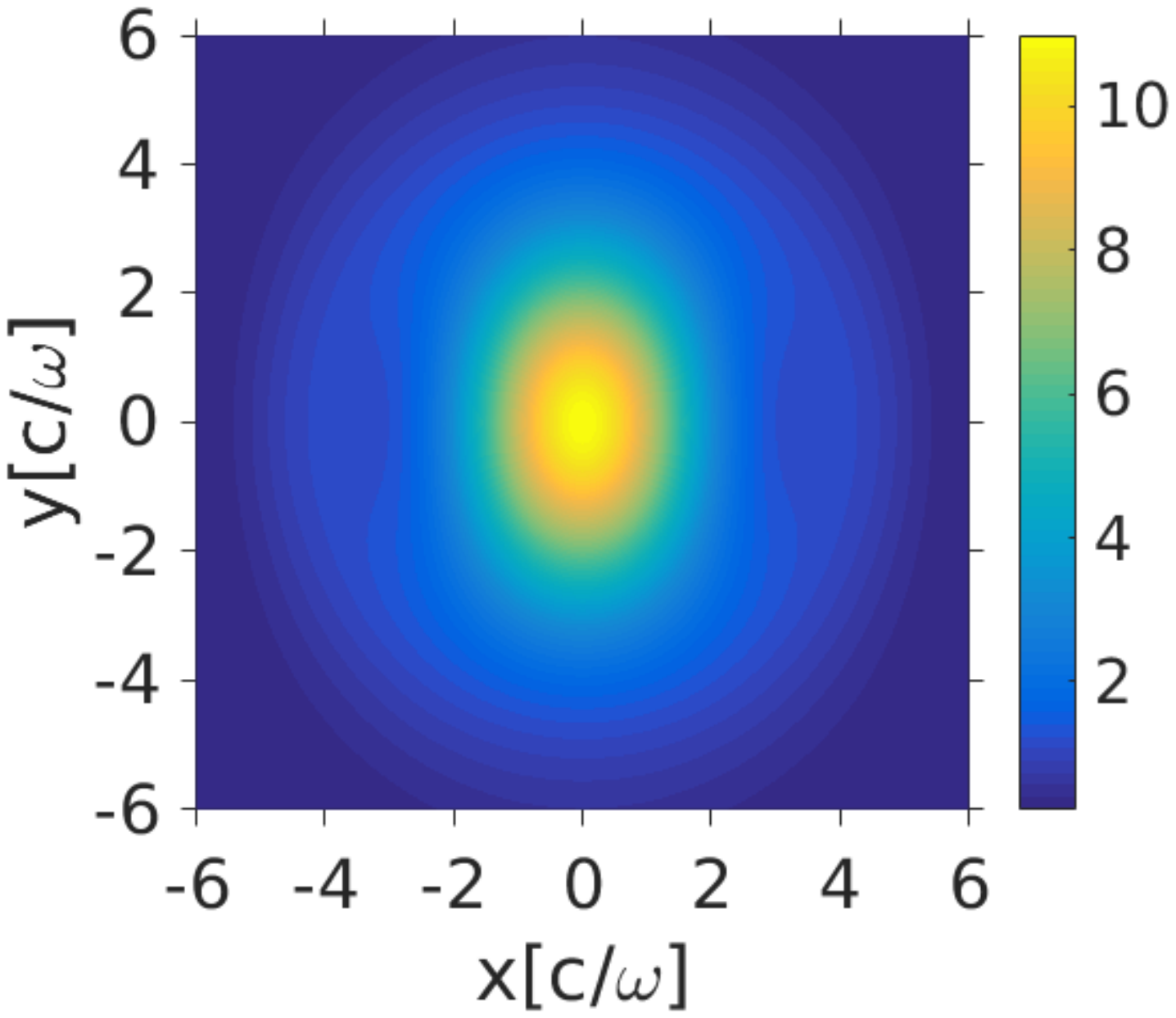}&
\includegraphics[width=0.335\textwidth,trim= 2cm 6cm 0cm 2cm,clip=true]{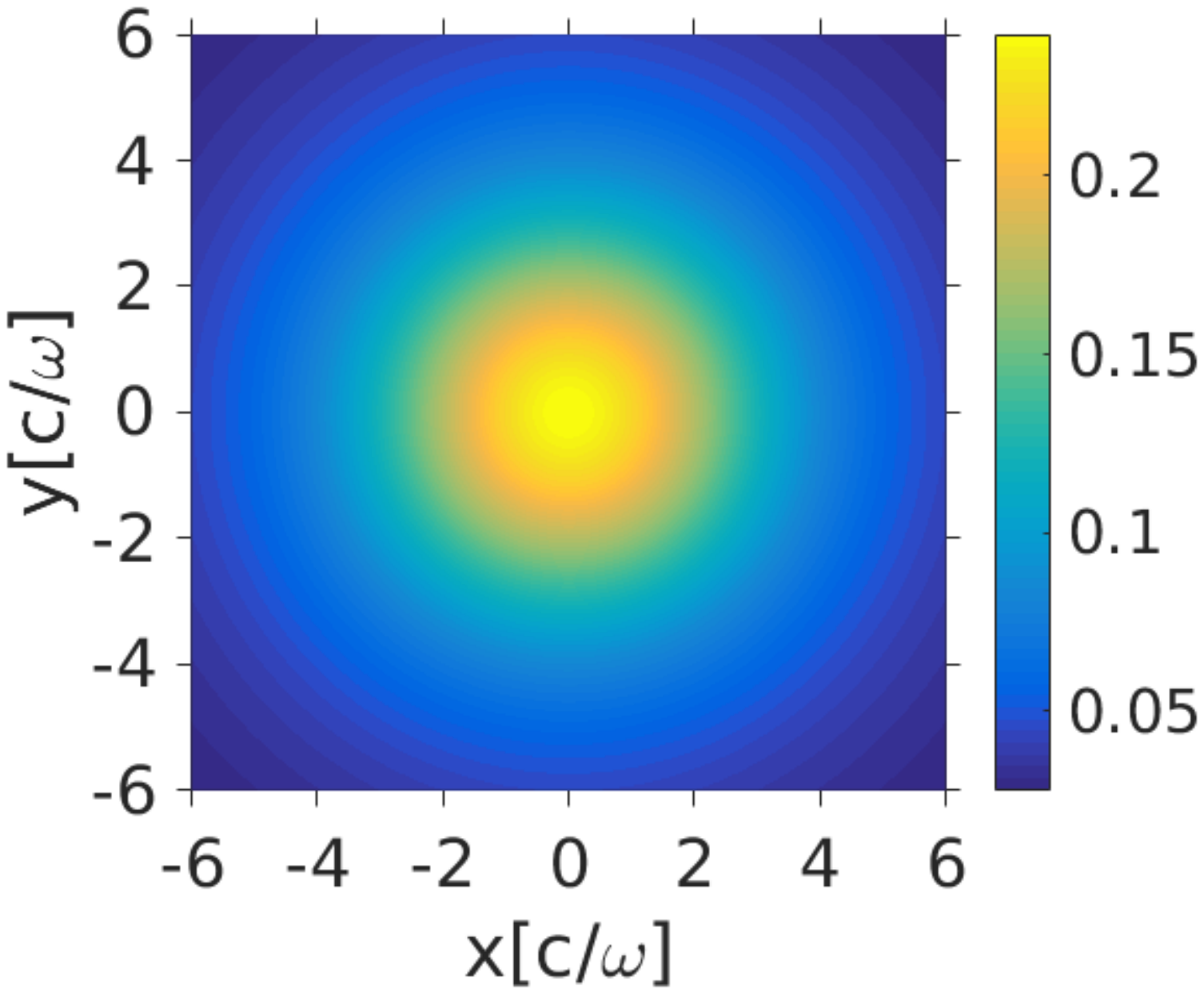}&
\includegraphics[width=0.335\textwidth,trim= 2cm 6cm 0cm 2cm,clip=true]{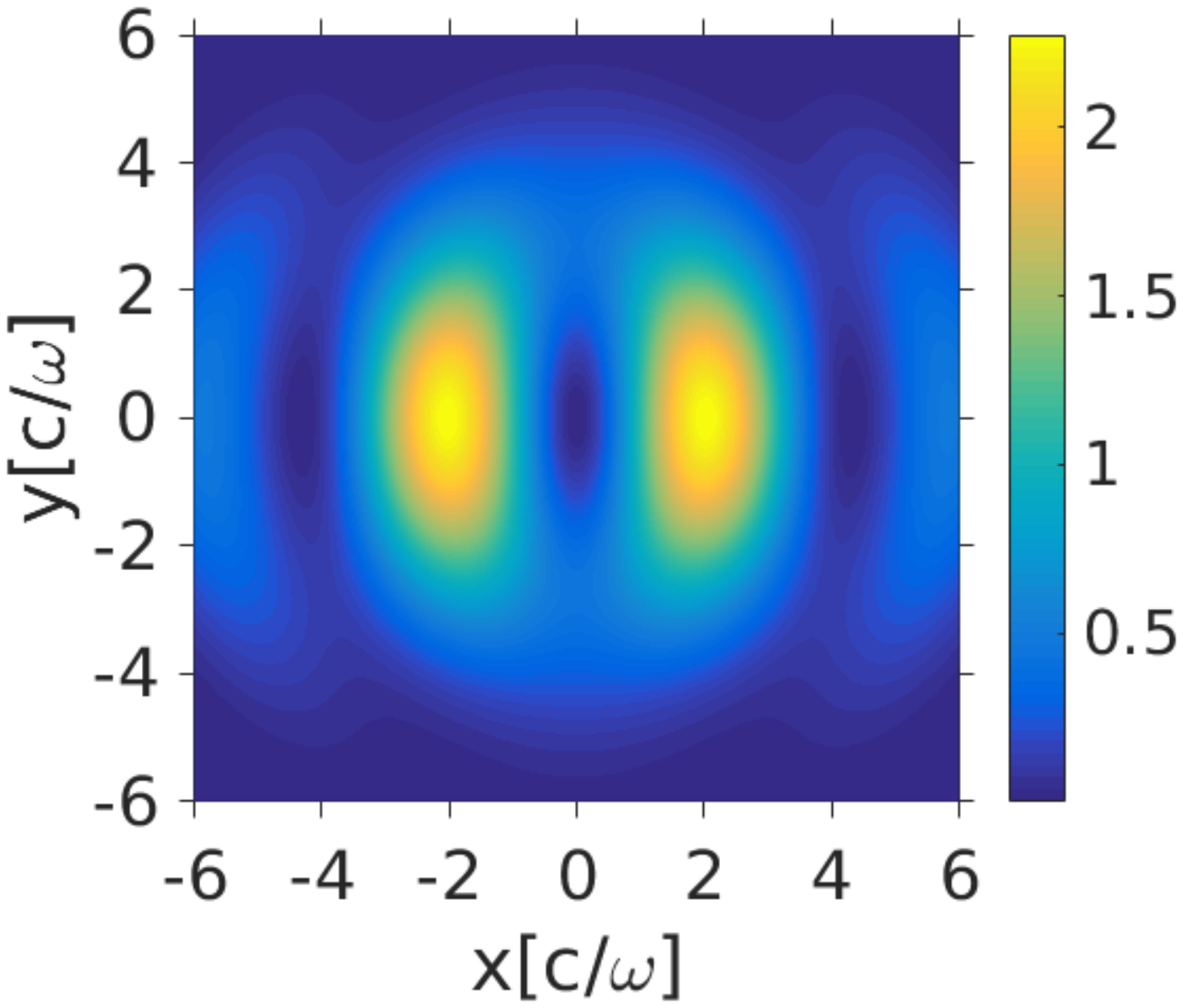}\\
 (d) & (e) & (f)\\
\end{tabular} 
\caption{ Illustrative example of the density of EM energy, Eq.(\ref{eq:energyden}), at the (a) plane defined by $x_2 =0$ and at the (b) plane defined by $x_3 =0$; (c) illustrates the electric field projection into  the plane $x_3 =0$; in (d)  $\vert \mathbf{E}\times\mathbf{e}_3\vert^2 $; (e) $\vert B_r\vert ^2$;  (f) $\vert \mathbf{B}\cdot\mathbf{e}_3\vert^2 $are shown at such  plane.
All subfigures refer to a Dirichlet mode with the parameter $\kappa= -.0557606$ and the coefficient $\tilde c_0 =1$;  this $\kappa$ is the root with smallest absolute value of the Dirichlet boundary condition, Eq.~(\ref{eq:diri}), for a mirror surface at $\zeta_0 = 35700 c/\omega$. }\label{fig:dir}
\end{figure}

\subsubsection{$\{\mathbf{E}_{\mathcal{B}},\mathbf{B}_{\mathcal{B}}\}$ symmetrized modes for $m>0$}

We now consider the general structure of the symmetrized Hertz potential given by Eq.~(\ref{eq:pispacea}-\ref{eq:pispacec}) in configuration space and by Eq.~(\ref{eq:piwavesa}-\ref{eq:piwavesd}) in wave-vector space, to construct the
elementary $\mathcal{B}$-modes for $m>0$. The boundary conditions for $\mathcal{B}$-modes, Eqs.~(\ref{eq:E_B1}) and~(\ref{eq:E_B2}), are equivalent to a set of linear equations for the coefficients
$\{\tilde c^{\pm},\tilde c_{\kappa\pm i/2}^0\}_{\mathcal{B}}$; in matricial form these equations are
$$\mathbb{M}_{\mathcal{B}}\mathbb{C}_{\mathcal{B}}=:$$\begin{equation}
\begin{pmatrix}
d_++d_-\mathcal{W}_{\kappa,m}&\!\mathcal{W}_{\kappa,m}-1  &\!-\mathcal{W}_{\kappa,m}&\!0\\	
1&\!\mathcal{W}_{\kappa,m} &-\mathcal{W}_{\kappa,m}-1&0\\
d_++d_-\mathcal{W}_{\kappa,m}&\!1-\mathcal{W}_{\kappa,m}  &\! 0 &\! 1\\
d_+& -d_- \mathcal{W}_{\kappa,m}&\!0 &\!d_+-d_-\mathcal{W}_{\kappa,m}
\end{pmatrix}
\begin{pmatrix}
\tilde c^+_{\kappa,m}\\
\tilde c^-_{\kappa,m}\\
\tilde c^0_{\kappa+i/2,m}\\
\tilde c^0_{\kappa-i/2,m}\end{pmatrix} 
= 0\label{eq:matB},
\end{equation}
with
\begin{equation}
\mathcal{W}_{\kappa,m} =\frac{V_{\kappa+i/2,m}(\zeta_0)}{V_{\kappa-i/2,m}(\zeta_0)}.
\end{equation}
The existence of  nontrivial solutions to this equation is conditioned to the existence of $\kappa$ values for which
\begin{equation}
\mathrm{Det}\mathbb{M}_{\mathcal{B}} = (d_++d_-\mathcal{W}_{\kappa,m}^2)(d_+\mathcal{W}_{\kappa,m}-d_-\mathcal{W}_{\kappa,m}^2 -d_+).
\end{equation}
Since $\vert d_+/d_-\vert = 1 = \vert  \mathcal{W}_{\kappa,m}\vert$, the condition
\begin{equation}
d_++d_-\mathcal{W}_{\kappa,m}^2 =0 \Rightarrow \mathcal{W}^2_{\kappa,m}=\frac{V^2_{\kappa+i/2,m}(\zeta_0)}{V^2_{\kappa-i/2,m}(\zeta_0)} = -\frac{d_+}{d_-}\label{eq:bc1}
\end{equation}
is feasible, while the condition
\begin{equation}
d_+\mathcal{W}_{\kappa,m}-d_-\mathcal{W}_{\kappa,m}^2 -d_+ = 0 \Rightarrow 1 =\vert \mathcal{W}_{\kappa,m} \vert =\frac{1}{2}\bigg| 1 \pm\sqrt{1 -4\frac{d_+}{d_-} } \bigg|
\end{equation}
is not.
Notice that 
\begin{equation}
\frac{d_-}{d_+}\mathcal{W}_{\kappa,m}+ \mathcal{W}_{\kappa,m}^*=0\quad\Rightarrow\quad d_-V^2_{\kappa+i/2,m}(\zeta_0) +d_+V^2_{\kappa-i/2,m}(\zeta_0) = 0,
\label{eq:bc}\end{equation} 
which involves just a real valued function on the left hand side, since $V_{\kappa+i/2,m}(\zeta_0)=V^*_{\kappa-i/2,m}(\zeta_0)$ for real
$\kappa$ and $\zeta_0$, Eq.~(\ref{eq:real}).

Two sets of solutions to Eq.~(\ref{eq:matB}) are
\begin{eqnarray}
\tilde c^+_{\kappa,m} &=& 0,\nonumber\\ \tilde c^0_{\kappa+i/2,m} &=& (1- \mathcal{W}_{\kappa,m}^*)d_+ \tilde c^-_{\kappa,m},\nonumber\\
\tilde c^0_{\kappa-i/2,m} &=&-(1- \mathcal{W}_{\kappa,m})d_- \tilde c^-_{\kappa,m}\label{eq:B1}
\end{eqnarray}
and
\begin{eqnarray}
\tilde c^-_{\kappa,m} &=& 0,\nonumber\\ \tilde c^0_{\kappa+i/2,m} &=& (d_+\mathcal{W}_{\kappa,m}^* +d_-) \tilde c^+_{\kappa,m},\nonumber\\
\tilde c^0_{\kappa-i/2,m} &=&-(d_-\mathcal{W}_{\kappa,m} +d_+) \tilde c^+_{\kappa,m}.\label{eq:B2}
\end{eqnarray}

Notice that $\mathcal{W}_{\kappa,m}^*=\mathcal{W}_{\kappa,m}^{-1}$ and the boundary condition Eq.(\ref{eq:bc1}) guarantees that the coefficients
$\{\tilde c^{\pm},\tilde c_{\kappa\pm i/2}^0\}_{\mathcal{B}}$ satisfy Eq.~(\ref{eq:acs}); as a consequence $\mathbf{E}_{\mathcal{B}}$
and $\mathbf{B}_{\mathcal{B}}$ are eigenvectors of $\hat J_3$ and $\hat{\mathfrak{A}}_3$. Note also that the two sets of
solutions, Eqs.~(\ref{eq:B1}) and~(\ref{eq:B2}), lead to the same electromagnetic mode since a linear combination of them can be written as a gradient of a field, Eq.~(\ref{eq:trivial}).

The $\mathcal{B}$-modes are illustrated for small values of $\kappa$ in  Figure \ref{fig:mb}. An optical vortex is observed for the $x_3$ component of the electric field  at the $XY$-plane.
Notice that the components of the electric field exemplified in Fig.~\ref{fig:mb}c correspond to the real part of Eq.~(\ref{eq:Bmodeb}), while Fig.~\ref{fig:mb}d shows the modulus $(\mathbf{E}\times\mathbf{e}_3)\cdot(\mathbf{E}^*\times\mathbf{e}_3)$. For the illustrated $\mathcal{B}$-mode,  the  EM energy density is of purely magnetic origin at the focus of the mirror.
\begin{figure}
\begin{tabular}{@{}c @{}c @{}c}
\includegraphics[width=0.33\textwidth,trim= 2cm 6cm 0cm 2cm,clip=true]{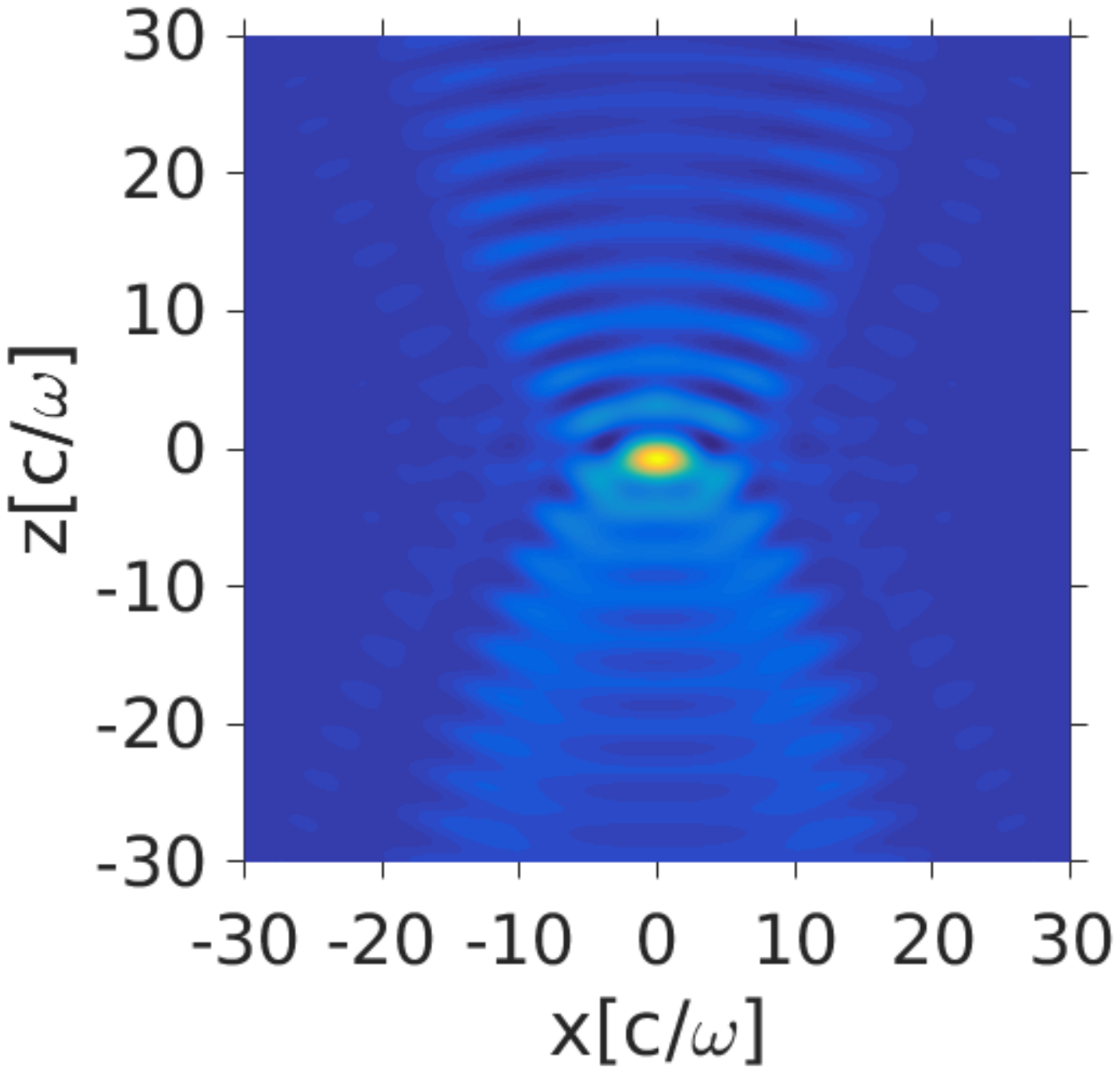}&
 \includegraphics[width=.33\textwidth,trim= 2cm 6cm 0cm 2cm,clip=true]{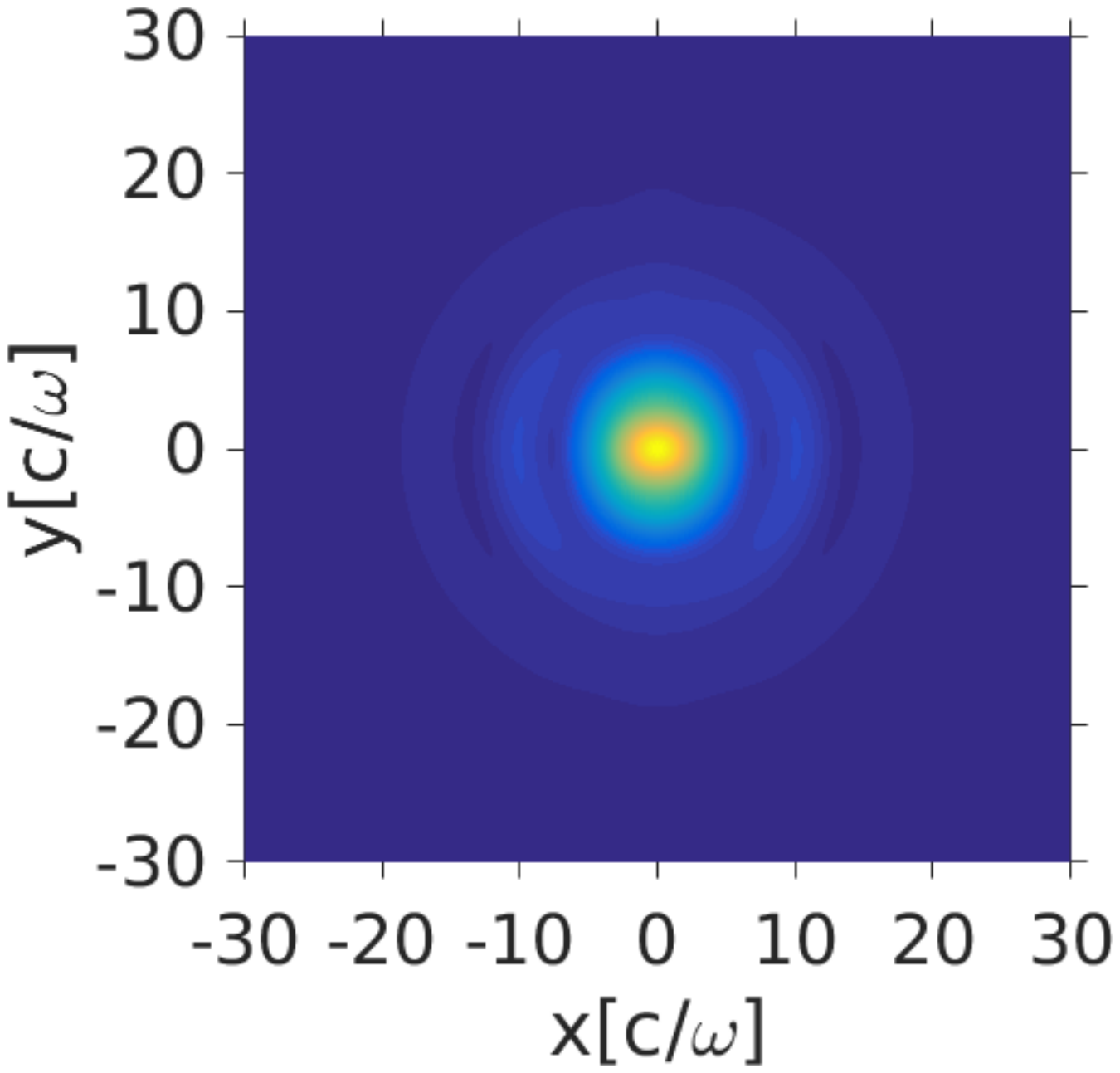}&
\includegraphics[width=0.33\textwidth,trim= 2cm 6cm 0cm 2cm,clip=true]{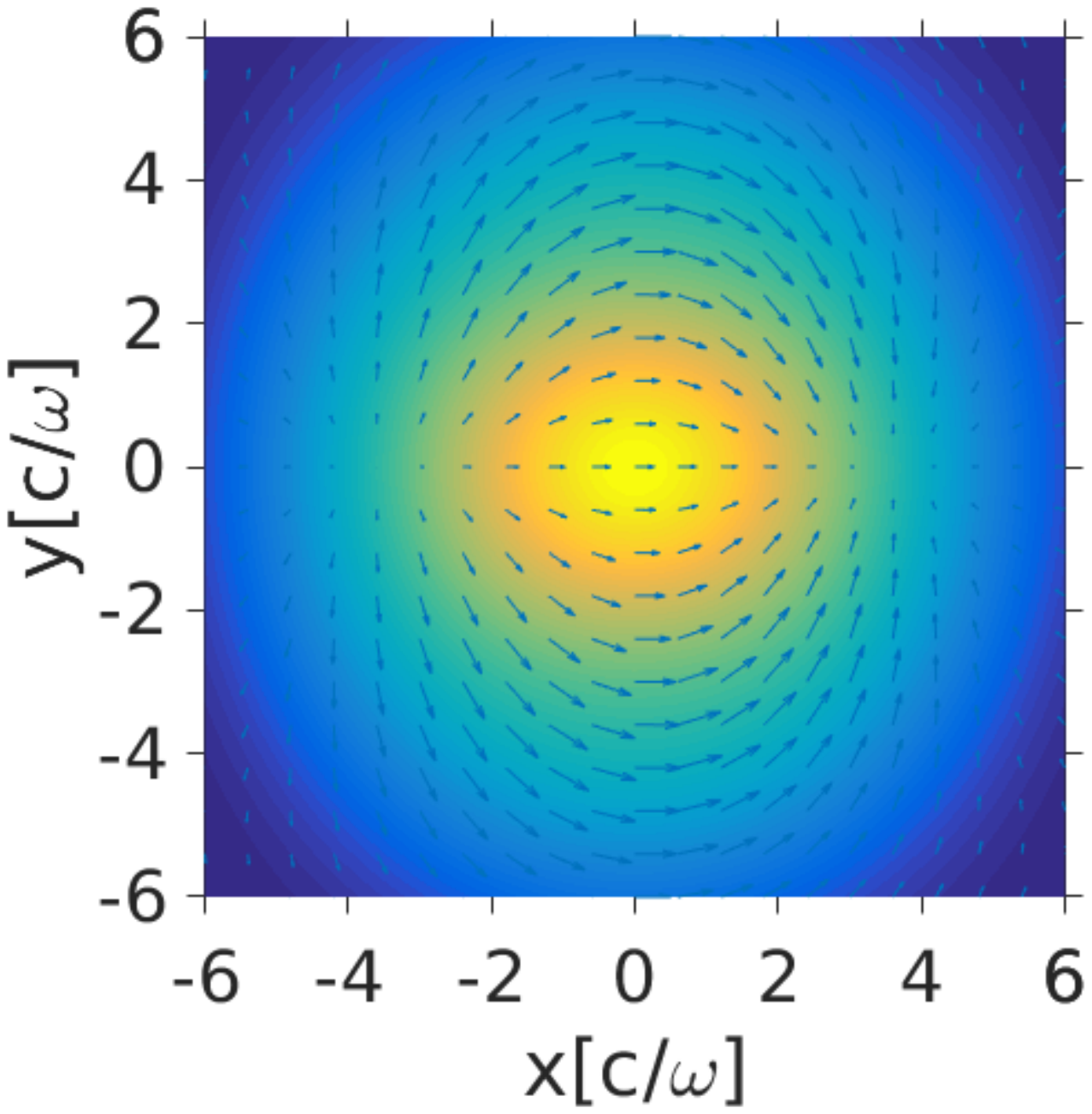}\\
(a) &(b) &(c) \\
\includegraphics[width=0.33\textwidth,trim= 2cm 6cm 0cm 2cm,clip=true]{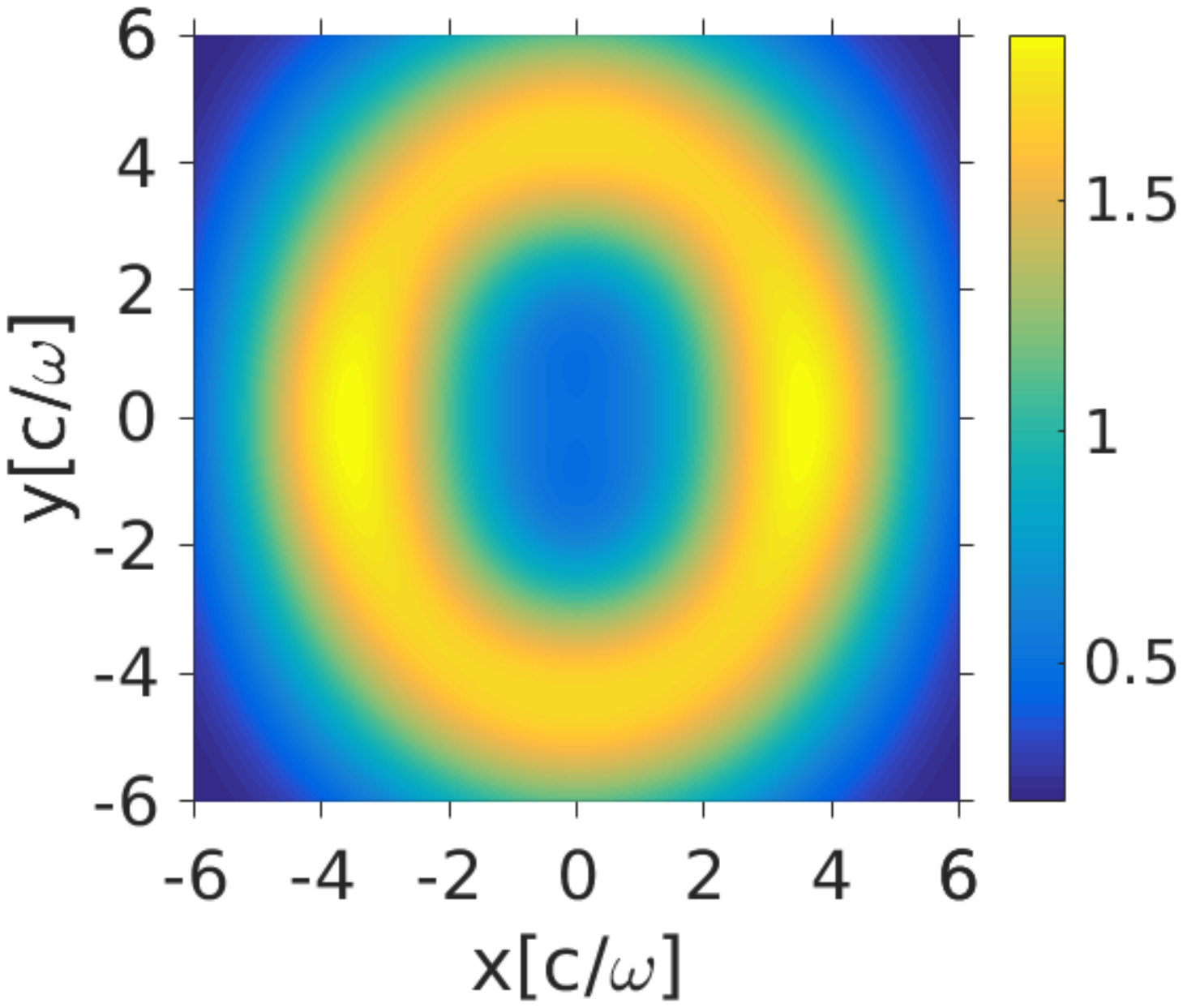}&
\includegraphics[width=0.33\textwidth,trim= 2cm 6cm 0cm 2cm,clip=true]{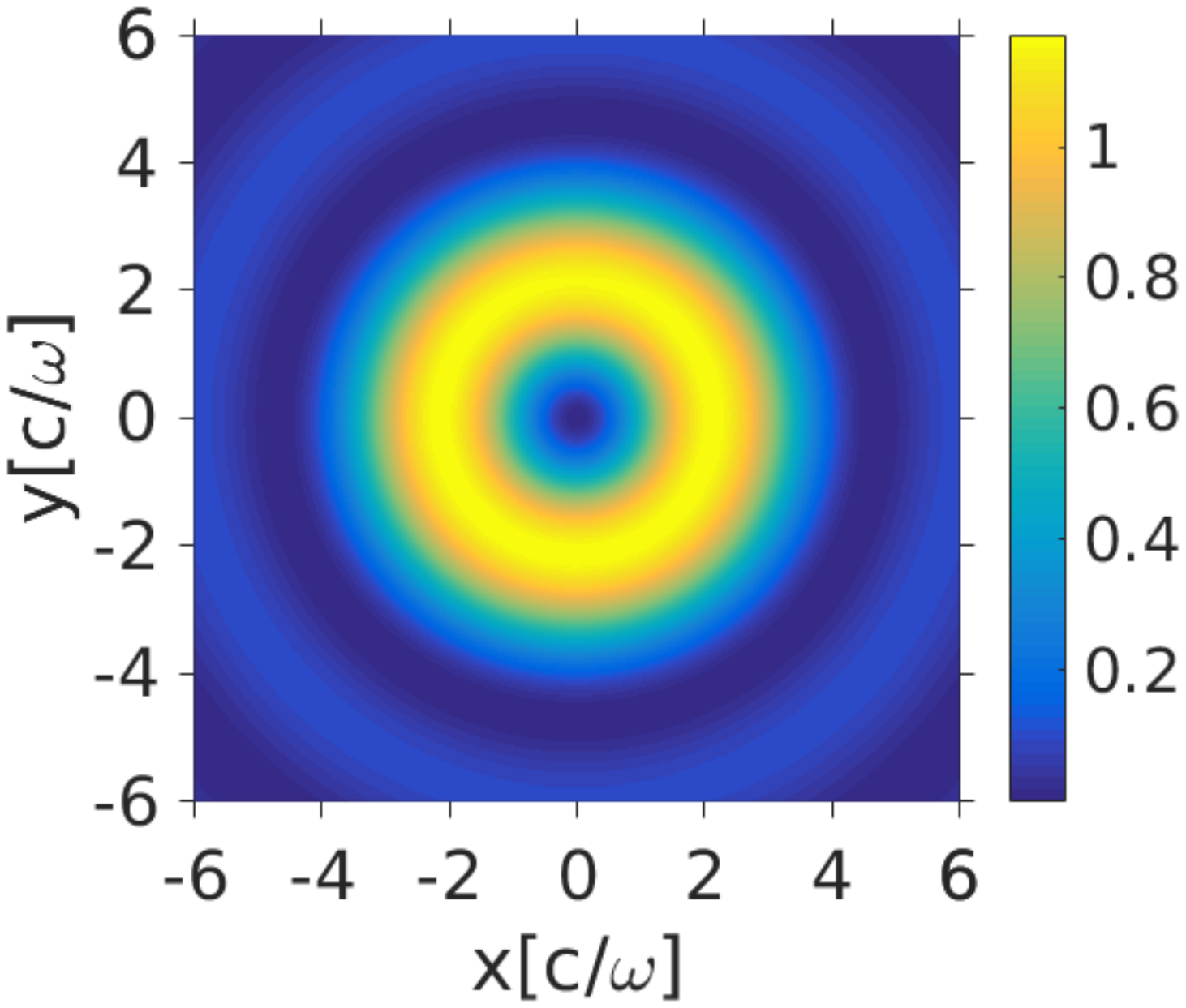}&
\includegraphics[width=0.33\textwidth,trim= 2cm 6cm 0cm 2cm,clip=true]{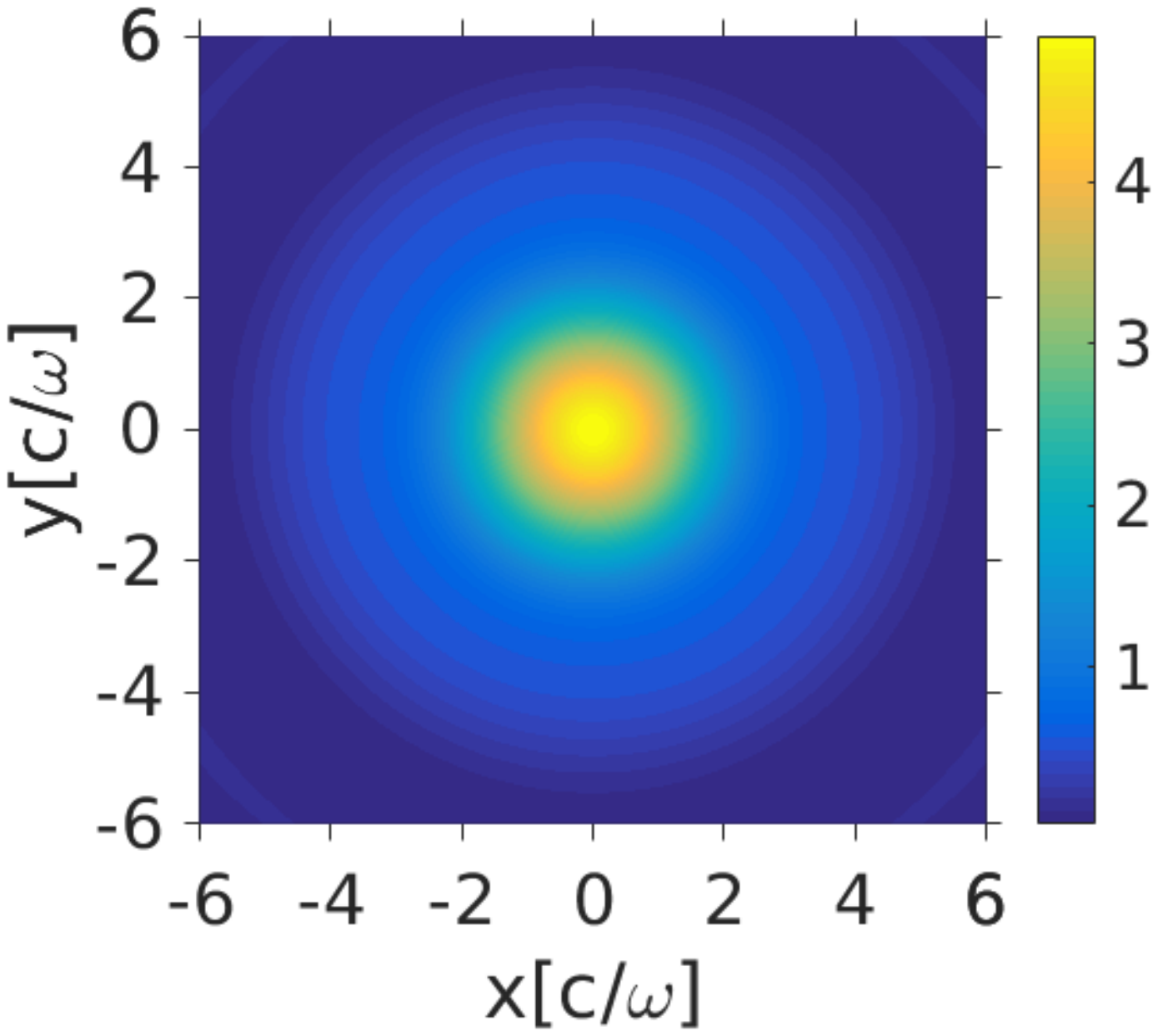}\\
 (d) & (e) & (f)\\
\end{tabular} 
\caption{ Illustrative example of the density of EM energy Eq.(\ref{eq:energyden}) at the (a) plane defined by $x_2 =0$ and at the (b) plane defined by $x_3 =0$; (c) illustrates the electric field projection into  the plane $x_3 =0$;(d)  $\vert {\mathbf{E}}\times\mathbf{e}_3\vert ^2$; (e) $\vert {\mathbf{E}}\cdot\mathbf{e}_3\vert ^2$;  (f) $\vert {\mathbf{B}}\times\mathbf{e}_3\vert ^2$  are also shown at the $x_3=0$  plane.
All subfigures refer to a $\mathcal{B}$-mode with $m=1$, $c^+_{\kappa,1} =1$, $c^-_{\kappa,1} =0$,  and the parameter $\kappa=0.03237277$;  this value of $\kappa$ has the smallest absolute value among the  roots  of the boundary condition Eq.~(\ref{eq:bc}) for a mirror surface at $\zeta_0 = 35700 c/\omega$. }\label{fig:mb}
\end{figure}

\subsubsection{$\{\mathbf{E}_{\mathcal{E}},\mathbf{B}_{\mathcal{E}}\}$ elementary modes for $m>0$}
The structure of a vectorial Hertz potential for a symmetrized EM mode is the same to that used for the $\{\mathbf{E}_{\mathcal{B}},\mathbf{B}_{\mathcal{B}}\}$ modes.
The boundary conditions for $\mathcal{E}$-modes, Eq.~(\ref{eq:E_E1}) and Eq.~(\ref{eq:E_E2}) lead to
the matricial equation
$$\mathbb{M}_{\mathcal{E}}\mathbb{C}_{\mathcal{E}}=:$$\begin{equation}
\begin{pmatrix}
1&\!- \mathcal{W}_{\kappa,m}  &\! \mathcal{W}_{\kappa,m}-1&\!0\\	
0&\!\mathcal{W}_{\kappa,m}+\frac{d_+}{d_-}\mathcal{W}^*_{\kappa,m}&-( \mathcal{W}_{\kappa,m}+\frac{d_+}{d_-}\mathcal{W}^*_{\kappa,m})&0\\
1&\!-\mathcal{W}^*_{\kappa,m}  &\! 0 &\! 1-\mathcal{W}^*_{\kappa,m}\\
0&\! \mathcal{W}^*_{\kappa,m}+\frac{d_-}{d_+}\mathcal{W}_{\kappa,m}&\!0 &\!\mathcal{W}^*_{\kappa,m}+\frac{d_-}{d_+}\mathcal{W}_{\kappa,m}
\end{pmatrix}
\begin{pmatrix}
\tilde c^+_{\kappa,m}\\
\tilde c^-_{\kappa,m}\\
\tilde c^0_{\kappa+i/2,m}\\
\tilde c^0_{\kappa-i/2,m}\end{pmatrix} 
= 0\label{eq:mat1},
\end{equation}
for the $\{\tilde c^{\pm},\tilde c_{\kappa\pm i/2}^0\}_{\mathcal{E}}$ in the wave vector representation of the Hertz potential.

The consistency of these equations requires,
\begin{equation}
\mathrm{Det}\mathbb{M}_{\mathcal{E}} = -2\left|\frac{d_-}{d_+}\mathcal{W}_{\kappa,m}+ \mathcal{W}_{\kappa,m}^*\right|^2=0.\label{eq:detE}
\end{equation}

A condition, involving a real valued function of $\zeta_0$, that {\it coincides with that for $\mathcal{ B}$ modes},
Eq.~(\ref{eq:bc}).

Assuming the fulfillment of Eq.~(\ref{eq:detE}), is equivalent to just two linear equations for
the four coefficients
\begin{eqnarray}
\tilde c^+_{\kappa,m}- \mathcal{W}_{\kappa,m}\tilde c^-_{\kappa,m}- (1- \mathcal{W}_{\kappa,m})\tilde c^0_{\kappa+i/2,m}&=&0,\\
\tilde c^+_{\kappa,m}- \mathcal{W}_{\kappa,m}^*\tilde c^-_{\kappa,m}+ (1- \mathcal{W}_{\kappa,m}^*)\tilde c^0_{\kappa-i/2,m}&=&0.
\end{eqnarray}
A complementary condition can be taken when symmetrized $\mathcal{E}$-modes are used. This
can be achieved in a similar way to the one used for $\mathcal{ B}$-modes. Two sets of
coefficients $\{\tilde c^{\pm},\tilde c_{\kappa\pm i/2}^0\}_{\mathcal{E}}$ that yield these conditions are $$\tilde c^+_{\kappa,m}=0,$$ with
\begin{eqnarray}
\tilde c^0_{\kappa+i/2,m}&=&\:\:\frac{1}{(\mathcal{W}_{\kappa,m}^*-1)}\tilde c^-_{\kappa,m},\\
\tilde c^0_{\kappa-i/2,m}&=&-\frac{1}{(\mathcal{W}_{\kappa,m}-1)}\tilde c^-_{\kappa,m} , \label{eq:set1E}
\end{eqnarray}
and $$\tilde c^-_{\kappa,m}=0,$$ with
\begin{eqnarray}
\tilde c^0_{\kappa+i/2,m}&=&-\frac{1}{(\mathcal{W}_{\kappa,m}-1)}\tilde c^+_{\kappa,m},\\
\tilde c^0_{\kappa-i/2,m}&=&\:\:\frac{1}{(\mathcal{W}_{\kappa,m}^*-1)}\tilde c^+_{\kappa,m}.\label{eq:set2E}
\end{eqnarray}
Both of them lead to the same set of electromagnetic modes: a linear combination of the Hertz potentials specified by Eqs.(\ref{eq:set1E}-\ref{eq:set2E}) can be found that satisfies Eq.~(\ref{eq:trivial}).

The $\mathcal{E}$- modes are illustrated in Figure \ref{fig:me}.  For the chosen $\mathcal{E}$-mode,  the  EM energy density is mostly of {\it electric} origin at the focus of the mirror. Three optical vortices are observed  for the $x_3$-component of the electric field at the $x_1$-axis in Fig~\ref{fig:me}e.}

\begin{figure}
\begin{tabular}{@{}c @{}c @{}c}
\includegraphics[width=0.335\textwidth,trim= 2cm 6cm 0cm 2cm,clip=true]{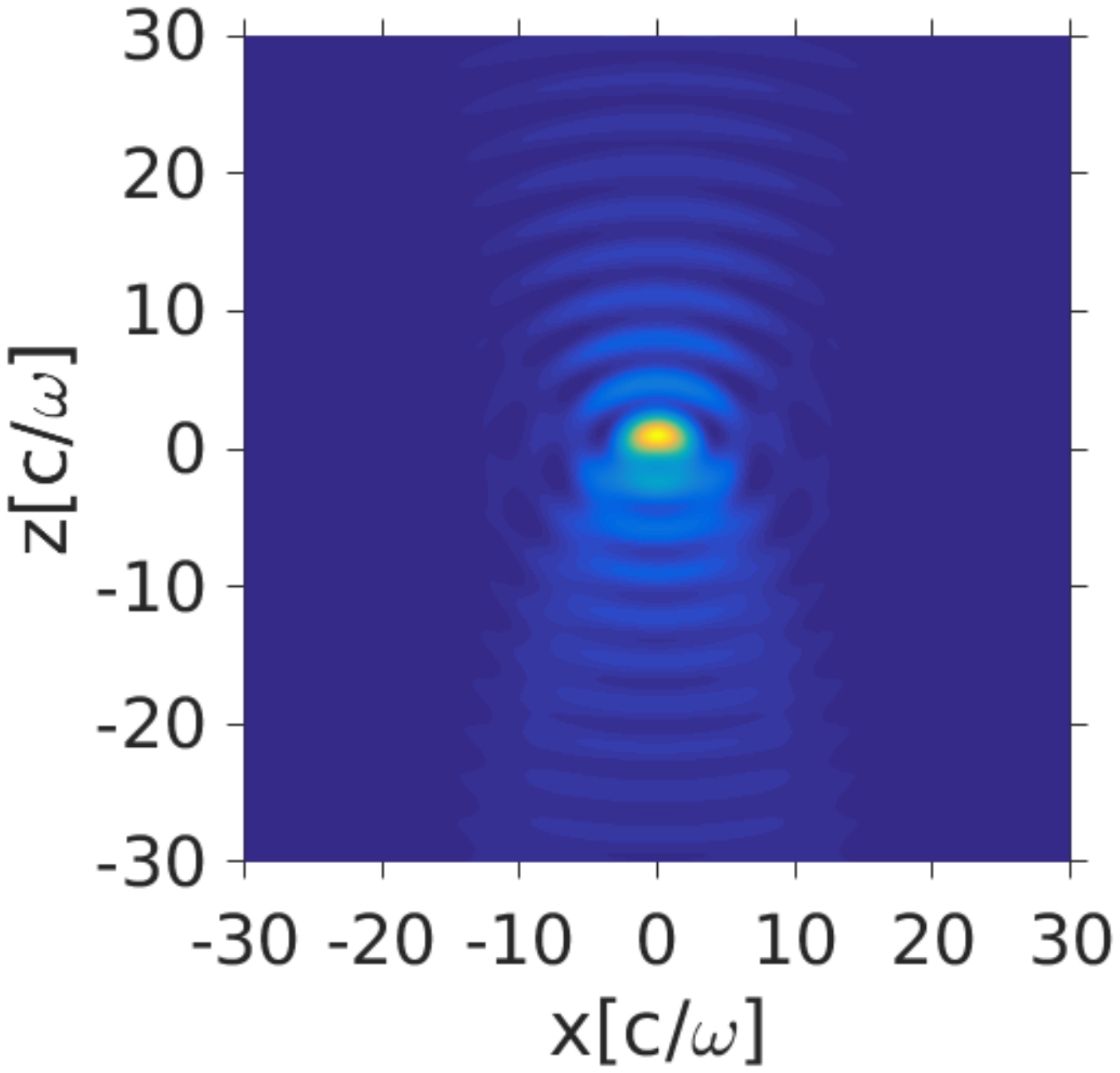}&
\includegraphics[width=0.335\textwidth,trim= 2cm 6cm 0cm 2cm,clip=true]{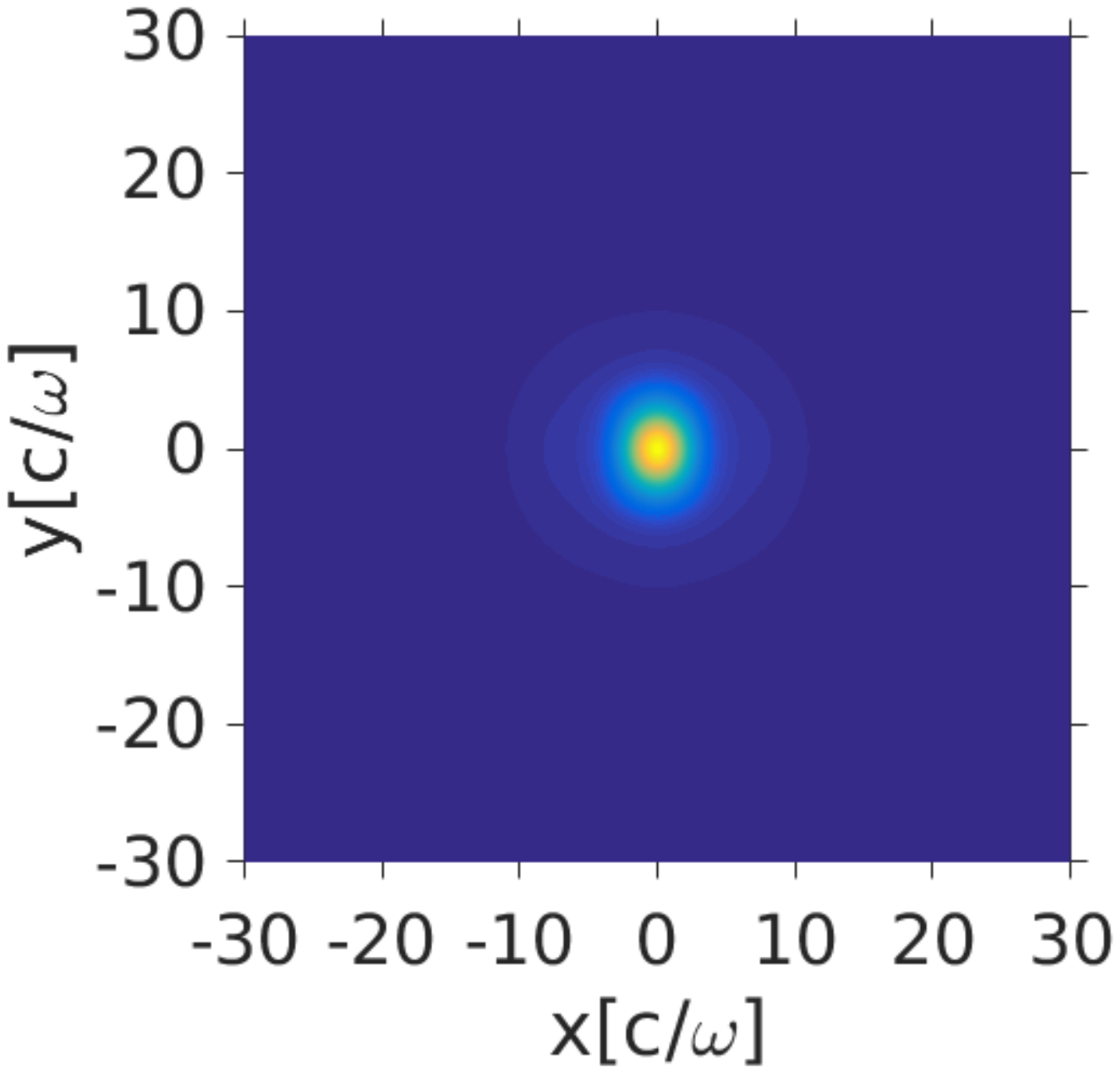}&
\includegraphics[width=0.335\textwidth,trim= 2cm 6cm 0cm 2cm,clip=true]{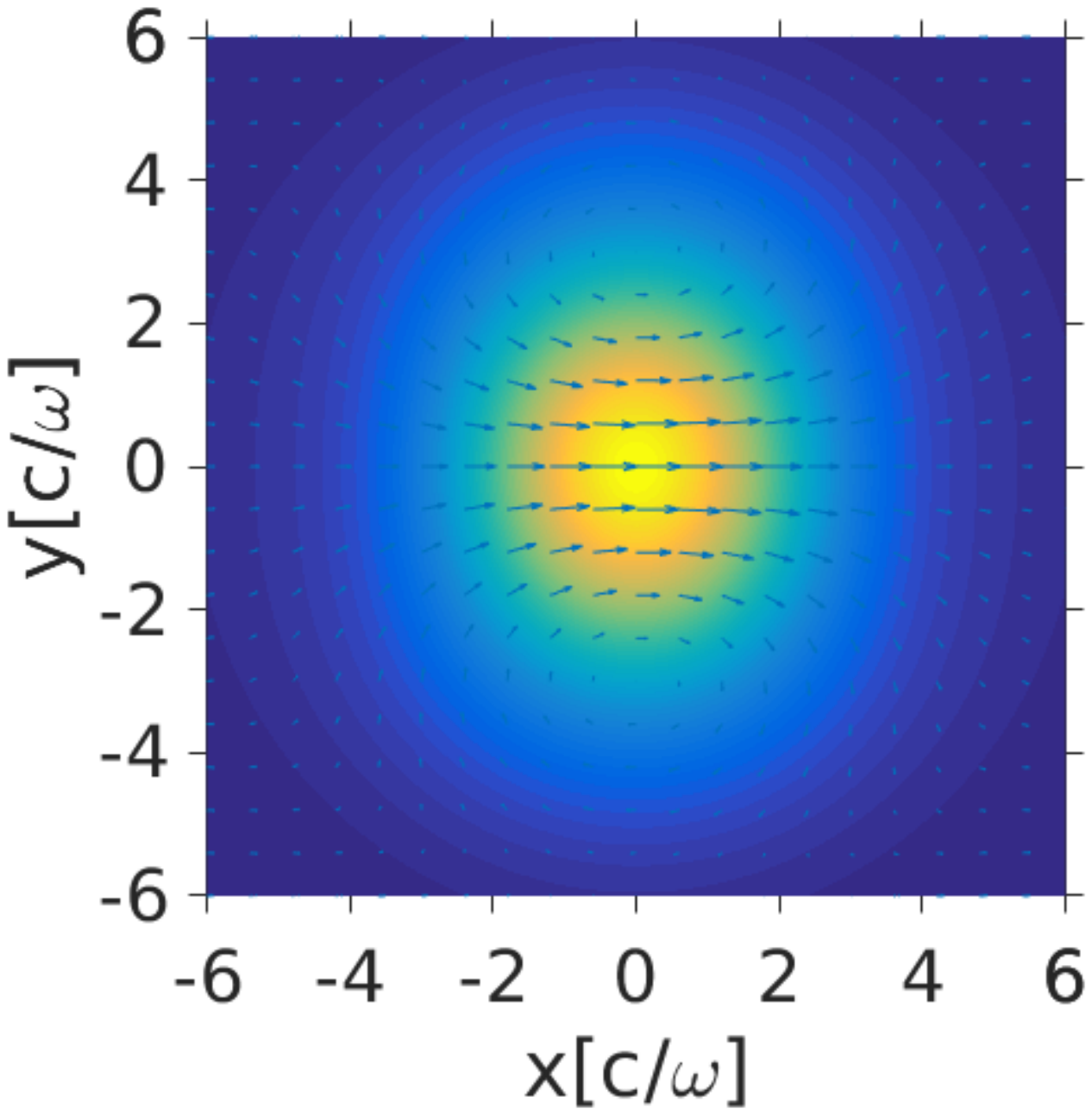}\\
(a) &(b) &(c) \\
\includegraphics[width=0.335\textwidth,trim= 2cm 6cm 0cm 2cm,clip=true]{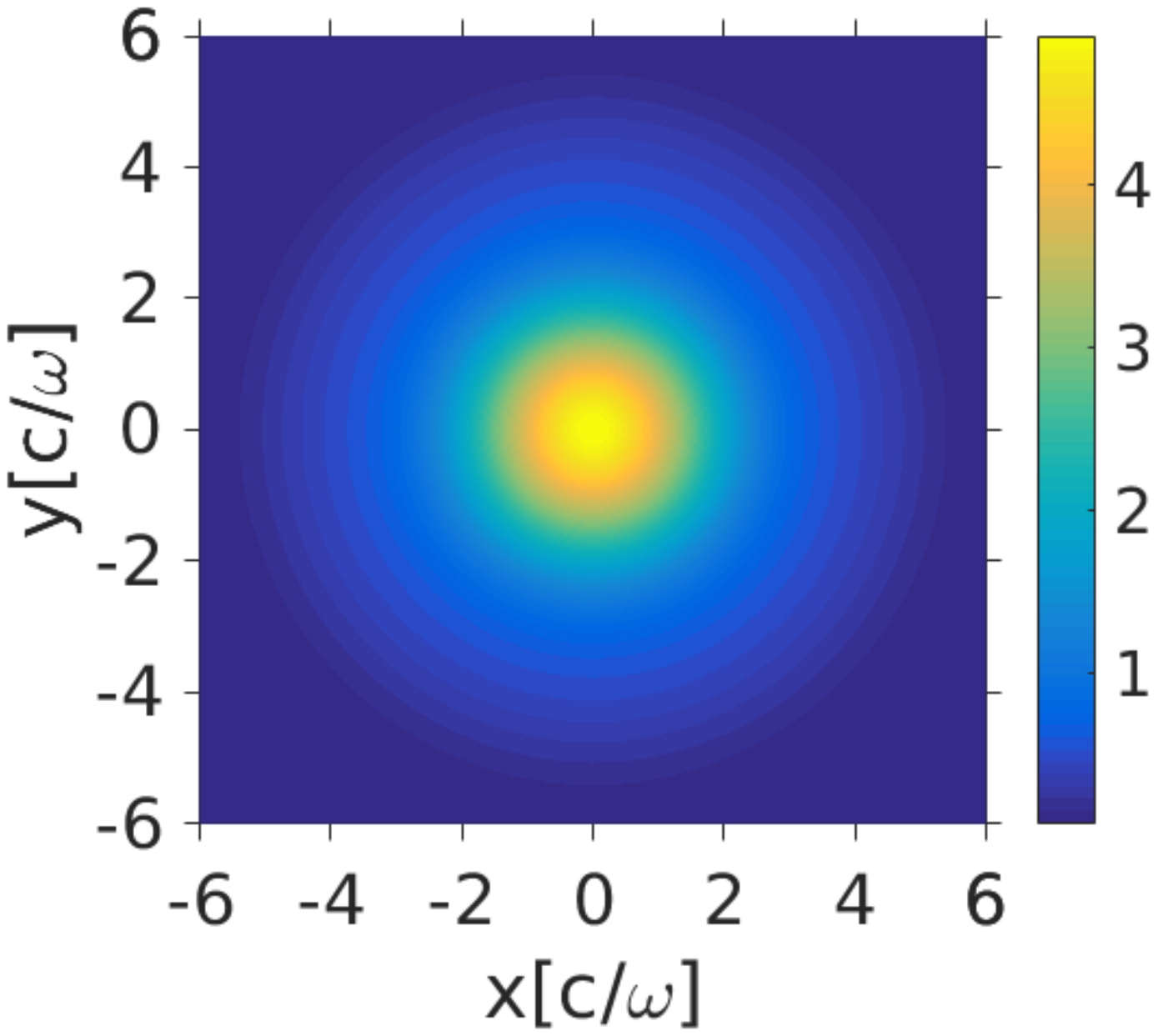}&
\includegraphics[width=0.335\textwidth,trim= 2cm 6cm 0cm 2cm,clip=true]{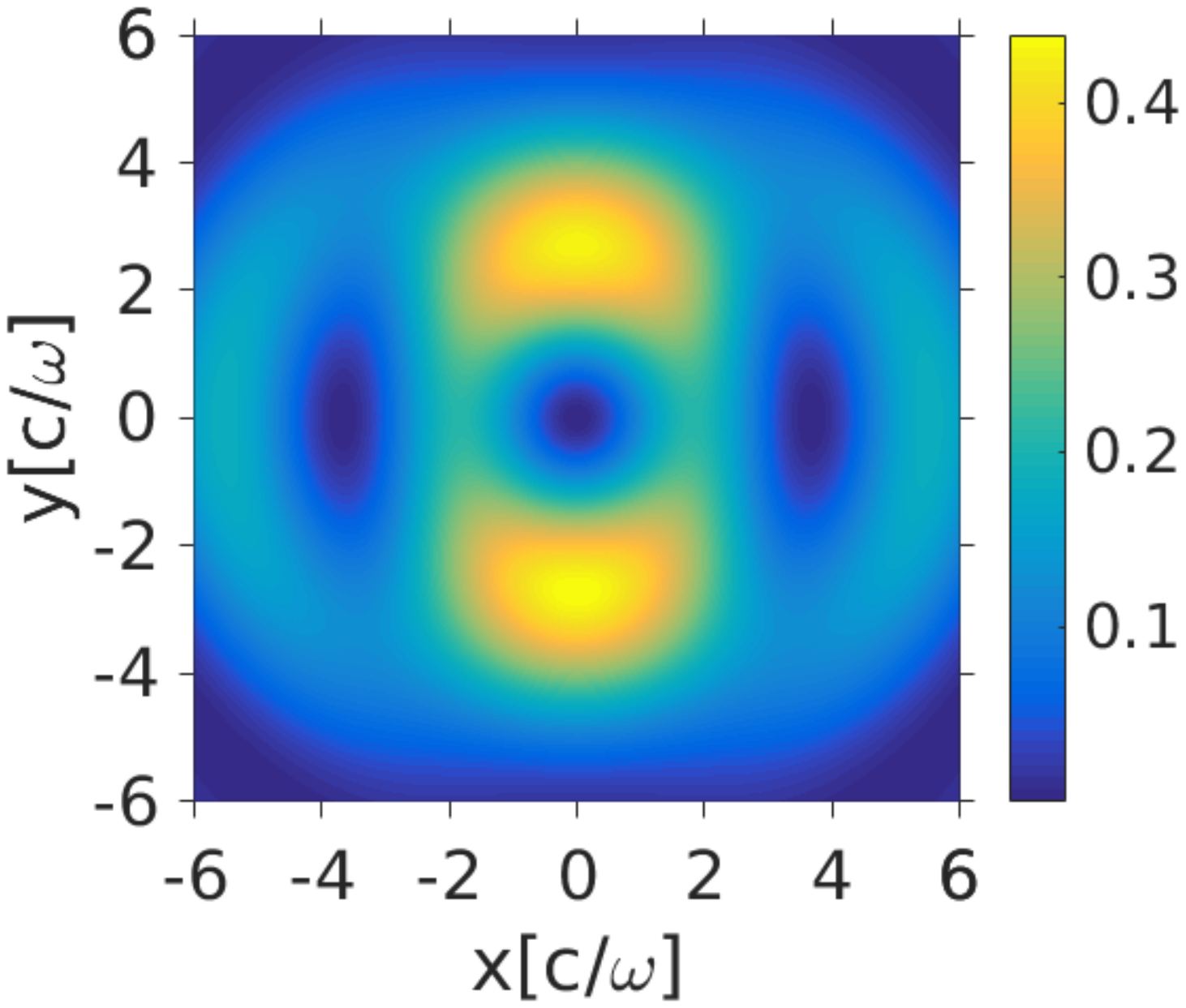}&
\includegraphics[width=0.335\textwidth,trim= 2cm 6cm 0cm 2cm,clip=true]{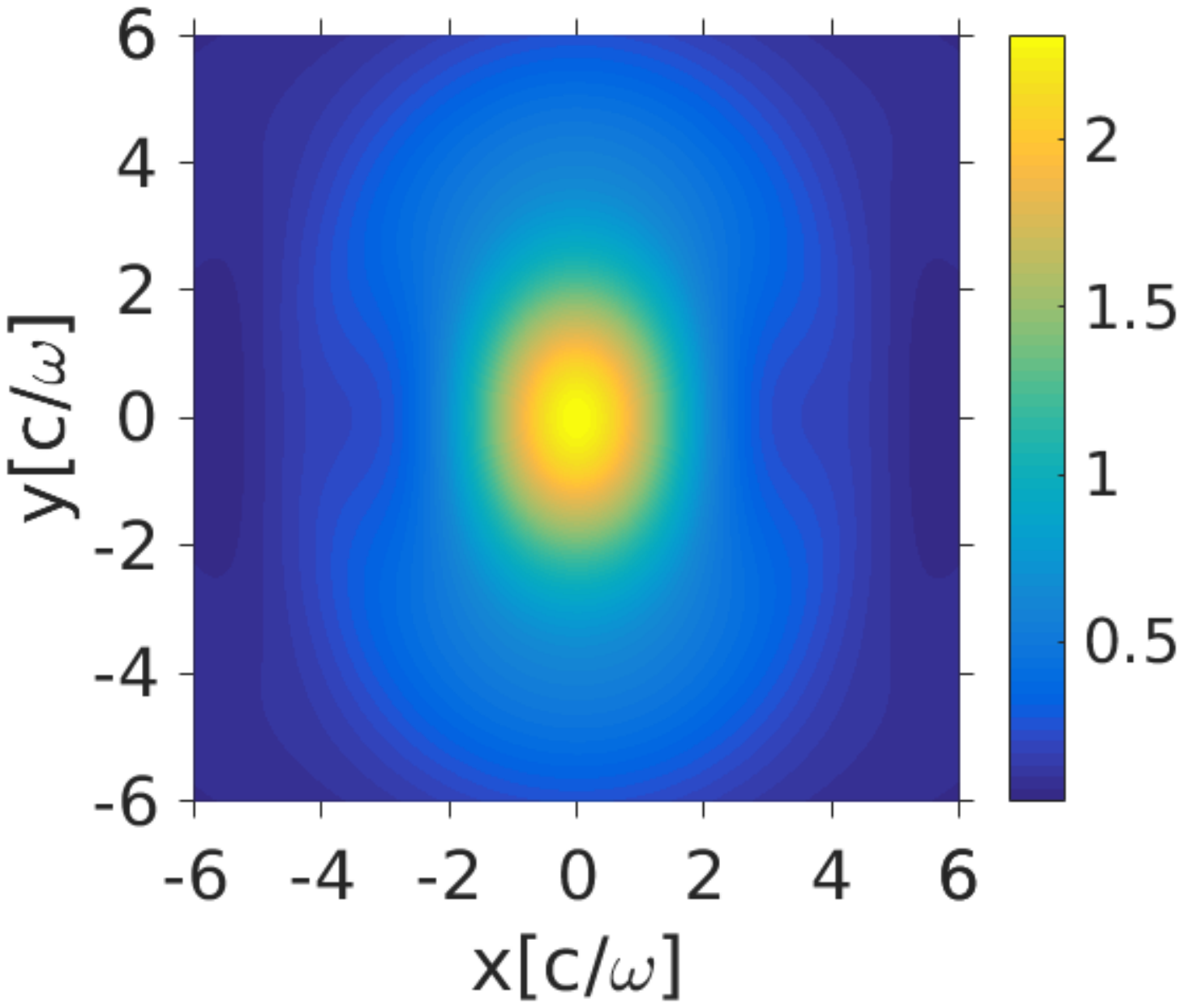}\\
 (d) & (e)&(f)\\
\end{tabular}
\caption{ Illustrative example of the density of EM energy Eq.(\ref{eq:energyden}) at the (a) plane defined by $x_2 =0$ and at the (b) plane defined by $x_3 =0$; (c) illustrates the electric field projection into  the plane $x_3 =0$;(d)  $\vert {\mathbf{E}}\times\mathbf{e}_3\vert ^2$; (e) $\vert {\mathbf{E}}\cdot\mathbf{e}_3\vert ^2$;  (f) $\vert {\mathbf{B}}\times\mathbf{e}_3\vert ^2$ 
are illustrated at the $x_3=0$  plane.
All subfigures refer to a $\mathcal{E}$-mode with $m=1$, $c^+_{\kappa,1} =1$, $c^-_{\kappa,1} =0$, and the parameter $\kappa=0.03237277$;  this  $\kappa$ has the smallest absolute value among the  roots  of the boundary condition Eq.~(\ref{eq:bc}) for a mirror surface at $\zeta_0 = 35700 c/\omega$. }\label{fig:me}
\end{figure}

\subsubsection{Elementary modes $\{\mathbf{E}_{\mathcal{B}},\mathbf{B}_{\mathcal{B}}\}$ and $\{\mathbf{E}_{\mathcal{E}},\mathbf{B}_{\mathcal{E}}\}$, $m< 0$}
The results shown in the previous two sections for the ${\mathcal{B}}$-modes and ${\mathcal{E}}$-modes  used  the identities Eqs.(\ref{eq:v1}-\ref{eq:v6}) which are valid for $m>0$.
For the $m<0$ modes, similar results can be obtained by noticing that the solutions of the wave equation for $m<0$ have a dependence on the $\zeta$ and $\eta$ variables, 
analogous to that for $m>0$ but interchanging the role of the functions  $V_{\kappa, m\pm 1}$ in the components $\pi_+$ and $\pi_-$ of the corresponding Hertz potential.

\section{Discussion}

We have shown that symmetries inherent in a parabolic geometry can be used to obtain closed expressions of the electromagnetic modes in presence of boundaries
exhibiting such geometry. Within the Hertz potentials formalism, these modes incorporate the vectorial character of the EM field into the two parameters given by
the elementary solutions of the scalar wave equation in a simple way. The intrinsic angular momentum---equivalent to a bivalued parameter $\sigma =\pm 1$---is imprinted on the form of the Hertz potentials: (i) it couples with the orbital angular momentum of the scalar field to yield a vector potential that only involves
wave functions with parameters $m$ and $m\pm 1$, (ii) it modifies the parabolic number $\kappa$ to yield symmetrized vectorial Hertz potentials via scalar wave functions with parameters $\kappa$ and $(2\kappa \pm 1\cdot i)/2$, both are shown in Eqs.~(\ref{eq:base_circ}).   

The elementary electromagnetic modes encountered from the Hertz potentials exhibit the underlying parabolic symmetry. They are orthonormal with respect to the scalar product defined by an extrapolation of the expression of the EM energy density for two different modes, Eq.~(\ref{eq:scalrp}). As a consequence, these modes can be used to directly perform Einstein quantization and define the quantum numbers of the corresponding photons. The quantum numbers are related to dynamical variables of the field and, in turn, its mechanical identity. The $m$ quantum number accounts for the projection of the total angular momentum along the $x_{3}$-axis, $J_{3}$, and the $\kappa$ quantum number is half the eigenvalue of the generator $\mathfrak{A}_{3}$ characteristic of parabolic symmetry. Both numbers could play a dynamical role in matter-EM field interactions, the study of which we leave for future work.

The relevance of using symmetrized modes has been further analized by working in detail the important case of an ideal parabolic mirror. We have shown that proper boundary conditions can be satisfied by these modes. For a mirror located at $\zeta_{0}$, the boundary conditions constrain the relative weights of the $\pi_\pm,\pi_0$ components of the Hertz potentials and limit the possible values of $\kappa$  to solutions of the compact expression $$\mathcal{W}^2_{\kappa,m}(\zeta_0) = -\frac{d_+}{d_-}, \quad d_\pm =\frac{1}{2} \pm i\frac{\kappa}{m},$$ which encloses the constrains found for both electric, Eqs.~(\ref{eq:neu},\ref{eq:detE}), and magnetic modes, Eqs.~(\ref{eq:dirichlet},\ref{eq:bc1}). We observed that, even though the boundary condition is naturally written in configuration space, several mathematical manipulations can be dealt in an easier way in the wave vector space.

The parabolic EM modes exhibit electric and magnetic fields with a non-trivial and rich topology.
Regions where electric and magnetic field have different magnitude, phase singularities, vectorial vortices, and strong gradients of the field components are found. These properties should be studied at depth in terms of their effect in the interaction with atomic systems.

The experimental generation of any given particular mode is an important subject. The most direct implementation corresponds to impringing the boundary conditions on the EM field through an accessible surface. In the optical realm the use of Space Light Modulators seems to be a promising option to that end. In fact this idea has already been studied and implemented for Neuman modes by Sonderman $et$ $al$ \cite{Sonder}. 

The results reported in this manuscript can also be used to study other interesting configurations, $e.g.$, cavities built from parabolic mirrors.
From the analysis reported here,  the complete set of modes inside a near confocal cavity bounded by the surfaces $\zeta =\zeta_0$ and $\eta =\eta_0$ is formed by fields with values of the parameter $\kappa$ that are simultaneos solutions of $$\mathcal{W}_{\kappa,m}^2(\zeta_0) = -\frac{d_+}{d_-},\quad\quad \mathcal{W}_{-\kappa,m}^2(\eta_0) = -\frac{d_-}{d_+}.$$ The feasibility of this condition should be studied in detail; including a comparison with  previous works that used the WKB approximation to analyse the EM field in confocal cavities \cite{noeckel}. 
A second example correspond to  explore the use the symmetrized vectorial EM fields  in the presence of parabolic lenses. 

{\bf Acknowledgements.} This work was partially supported by CONACyT LN-293471. R.J. thanks Gerd Leuchs for interesting conversations and Luis S\'anchez-Soto for stimulating discussions.

\begin{center}
{\bf Appendix A.} { Parabolic coordinates and the associated scale factors.}
\end{center}

 The parabolic coordinates are defined by:
\begin{equation}
x_1=\sqrt{\zeta\eta}\cos\varphi,\quad\quad x_2=\sqrt{\zeta\eta}\sin\varphi,\quad\quad x_3=\frac{1}{2}(\zeta - \eta)
\end{equation}
with
\begin{equation}
0\le\zeta<\infty,\quad\quad 0\le\eta<\infty,\quad\quad 0\le\varphi<2\pi.
\end{equation}
The scale factors are:
\begin{equation}
h_\zeta =\frac{1}{2}\sqrt{\frac{\zeta+\eta}{\zeta}},\quad\quad h_\eta =\frac{1}{2}\sqrt{\frac{\zeta+\eta}{\eta}},\quad\quad h_\varphi = \sqrt{\zeta\eta}.
\end{equation}
The unitary parabolic vectors are:
\begin{eqnarray}
\mathbf{e}_\zeta &=&\sqrt{\frac{\zeta}{\zeta+\eta}}\Big[\sqrt{\frac{\eta}{\zeta}}(\cos\varphi {\mathbf e}_1 +\sin\varphi {\mathbf e}_2) +{\mathbf e}_3\Big],\\
\mathbf{e}_\eta &=&\sqrt{\frac{\eta}{\zeta+\eta}}\Big[\sqrt{\frac{\zeta}{\eta}}(\cos\varphi {\mathbf e}_1 +\sin\varphi {\mathbf e}_2) -{\mathbf e}_3\Big],\\
\mathbf{e}_\varphi &=& -\sin\varphi {\mathbf e}_1 + \cos\varphi {\mathbf e}_2.\label{eq:paravec}
\end{eqnarray}
Inverting them one obtains the Cartesian basis:
\begin{eqnarray}
{\mathbf e}_1 &=& \frac{\sqrt{\eta}}{\sqrt{\zeta + \eta}}\cos\varphi {\mathbf e}_\zeta+
\frac{\sqrt{\zeta}}{\sqrt{\zeta + \eta}}\sin\varphi {\mathbf e}_\eta-
\sin\varphi {\bf e}_\varphi,\\
{\mathbf e}_2 &=& \frac{\sqrt{\eta}}{\sqrt{\zeta + \eta}}\sin\varphi {\mathbf e}_\zeta+
\frac{\sqrt{\zeta}}{\sqrt{\zeta + \eta}}\sin\varphi {\mathbf e}_\eta+
\cos\varphi {\mathbf e}_\varphi,\\
{\mathbf e}_3 &=& \frac{\sqrt{\zeta}}{\sqrt{\zeta + \eta}} \quad\quad{\mathbf e}_\zeta-
\frac{\sqrt{\eta}}{\sqrt{\zeta + \eta}} \quad\quad{\mathbf e}_\eta.
\end{eqnarray}
Notice that
\begin{equation}
\mathbf{e}_{\pm} = {\mathbf{e}}_1 \pm i {\mathbf{e}}_2 =\frac{e^ {\pm i\varphi}}{2}
\Big[ \frac{{\mathbf{e}}_\zeta}{h_\eta}
     +\frac{{\mathbf{e}}_\eta }{h_\zeta}\pm 2i{\mathbf{e}}_\varphi\Big].
\end{equation}

\begin{center}
{\bf Appendix B.} {$\mathcal{E}$-modes in the presence of an ideal parabolic mirror using partially symmetrized Hertz potentials}
\end{center}
\begin{center}
{\it $\{\mathbf{E}_{\mathcal{E}},\mathbf{B}_{\mathcal{E}}\}$ modes with $m=0$}
\end{center}
In this case $$\pi_{\pm}=\sum_\kappa c_{\kappa_0}^{(\pm)}
e^{\mp i\varphi}V_{\kappa,1}(\zeta)V_{-\kappa,1}(\eta)$$ so that
$$\pi_{\pm}=\sum_\kappa d_{\kappa_0}^{(\pm)}e^{\mp i\varphi}V_{-\kappa,1}(\eta)$$
with
$$d^{\pm}_{\kappa,0} = \tilde{c}_{\kappa +i/2,0}^{\pm}V_{\kappa +i/2,0} +\tilde{c}_{\kappa -i/2,0}^{\pm}V_{\kappa -i/2,0}.$$
Writing
$$\pi_0 = \sum_{\kappa} c_{\kappa,0}^{0}V_{\kappa,0}(\zeta)V_{-\kappa,0}(\eta) = \sum_\kappa d_{\kappa,0}^0V_{-\kappa,0}(\eta),\quad\quad d_{\kappa,0}^0 = c_{\kappa,0}^0V_{\kappa,0}(\zeta),$$
it can be shown that the condition $p_\eta = 0$ for $\zeta = \zeta_0$
is satisfied if
\begin{equation}
\sqrt{\zeta}\frac{\partial}{\partial\zeta}[d_{\kappa,0}^+ - d_{\kappa,0}^-] = i \sqrt{\zeta}\frac{\partial}{\partial\zeta}\Big[(-c_{\kappa+i/2}^+ + c_{\kappa+i/2}^-)V_{\kappa+i/2,1}(\zeta) +(c_{\kappa-i/2}^+ - c_{\kappa-i/2}^-)V_{\kappa-i/2,1}(\zeta)\Big].
\end{equation}
The equation $p_\varphi = 0$ is satisfied if
\begin{equation}
\sum_{\kappa}[\mu_{\kappa+i}^+ + \mu_\kappa^0 + \mu_{\kappa-i}^-]V_{-\kappa,m}(\eta) =0,
\end{equation}
with
\begin{eqnarray}
\mu_{\kappa+i}^+ &=& \Big(\frac{1}{2} + i\kappa -1\Big)\Big[
-\frac{1}{2\sqrt{\zeta}}(d_{\kappa+i}^+ + d_{\kappa+i}^-) - d_{\kappa+i,0}^0 - i\frac{\partial d_{\kappa+i}^0}{\partial\zeta}\Big],\nonumber\\
\mu_{\kappa}^0 &=&\frac{\partial}{\partial\zeta}\sqrt{\zeta}(d_{\kappa}^+ + d_\kappa^-) + \frac{1}{2\sqrt{\zeta}}(d_\kappa^+ + d_\kappa^-) + d_\kappa^0 - 2\kappa d_\kappa^0,\nonumber\\
\mu_{\kappa-i}^- &=& \Big(\frac{1}{2} - i\kappa -1\Big)\Big[-\frac{1}{2\sqrt{\zeta}}(d_{\kappa-i}^+ + d_{\kappa-i}^-) - d_{\kappa-i,0}^0 + i\frac{\partial d_{\kappa+i}^0}{\partial\zeta}\Big].\nonumber
\end{eqnarray}

\begin{center}
{\it $\{\mathbf{E}_{\mathcal{E}},\mathbf{B}_{\mathcal{E}}\}$modes, $m> 0$}
\end{center}
The vectorial Hertz potentials with well defined total angular momentum have the structure
\begin{eqnarray}
\tilde{\boldsymbol{\pi}}_\pm &=& e^{\pm i\varphi}\pi_\pm =e^{im\varphi}\sum_\kappa c^\pm_\kappa V_{\kappa,m\mp1}(\zeta)
V_{-\kappa,m\mp 1}(\eta),\nonumber\\
\pi_0 &=& e^{im\varphi}\sum_\kappa c^0_\kappa V_{\kappa,m}(\zeta) V_{-\kappa,m}(\eta).
\end{eqnarray}
Using the recurrence relations given above, it is possible to write
the functions $V_{-\kappa,m\pm 1}(\eta)$ in terms of functions $V_{-\kappa,m}(\eta)$, so that:
\begin{eqnarray}
P_+ &=& \frac{e^{im\varphi}}{2\sqrt{\eta}}\sum_\kappa (D^+_{\kappa +i/2}(\zeta)V_{-\kappa+i/2,m}(\eta) +D^+_{\kappa -i/2}(\zeta)V_{-\kappa-i/2,m})(\eta),\nonumber\\
P_- &=& \frac{e^{im\varphi}}{\sqrt{\eta}}\sum_\kappa (D^-_{\kappa +i/2}(\zeta)V_{-\kappa+i/2,m}(\eta) +D^-_{\kappa -i/2}(\zeta)V_{-\kappa-i/2,m})(\eta),
\end{eqnarray}
with
\begin{eqnarray}
D^+_{\kappa \pm i/2}(\zeta)&=& c^+_\kappa V_{\kappa,m-1}(\zeta)\Big(-\frac{1}{2} \mp \frac{i\kappa}{m}\Big) + c^-_\kappa V_{\kappa,m+1}(\zeta)(\mp i(\vert m\vert + 1)),\nonumber\\
D^-_{\kappa \pm i/2}(\zeta)&=& c^+_\kappa V_{\kappa,m-1}(\zeta)\Big(-\frac{1}{2} \mp \frac{i\kappa}{m}\Big) - c^-_\kappa V_{\kappa,m+1}(\zeta)(\mp i(\vert m\vert + 1)).
\end{eqnarray}
We can also write directly
\begin{eqnarray}
\pi_0 &=& e^{im\varphi}\sum_\kappa \tilde{c}^0_{\kappa + i/2}V_{-\kappa-i/2,m}(\eta) +
\tilde{c}^0_{\kappa - i/2}V_{-\kappa+i/2,m}(\eta), \nonumber\\
\tilde{c}^0_{\kappa \pm i/2}&=& c^0_{\kappa \pm i/2}V_{\kappa\pm i/2}(\zeta).
\end{eqnarray}
In terms of these factors, the condition $p_\eta =0$ at $\zeta=\zeta_0$ is satisfied
if
\begin{equation}
\tilde{c}^0_{\kappa \mp i/2} = -\frac{1}{\zeta} D^+_{\kappa\pm i/2} + \frac{2\sqrt{\zeta}}{m}
\frac{\partial}{\partial \zeta} D^-_{\kappa\pm i/2} , \quad \zeta = \zeta_0.
\end{equation}

The condition $p_\varphi = 0$ at $\zeta =\zeta_0$ is satisfied if:
$$\frac{\partial}{\partial\zeta}\sqrt{\frac{\zeta}{\eta}}P_+
 -\frac{\partial}{\partial\eta}\sqrt{\frac{\eta}{\zeta}}P_+= \Big[ \frac{\partial}{\partial \eta} + \frac{\partial}{\partial \zeta}\Big]\frac{\pi_0}{2}.$$
The expressions for the derivatives and products of the functions $V_{\kappa,m}$ by its argument lets write this equation in the form:
\begin{eqnarray}
\sum_\kappa (\lambda^{3/2}_{-\kappa+3i/2}V_{-\kappa+3i/2,m} &+&\lambda^{1/2}_{-\kappa+i/2}V_{-\kappa+i/2,m}
+\lambda^{-1/2}_{-\kappa-i/2}V_{-\kappa-i/2,m}\nonumber\\
&+&\lambda^{-3/2}_{-\kappa-3i/2}V_{-\kappa-3i/2,m})= 0
\end{eqnarray}
with
\begin{eqnarray}
\lambda^{3/2}_{-\kappa+3i/2} &=& (-i\tilde{c}^0_{\kappa -i/2} + \frac{\sqrt{\zeta}}{2}
\frac{\partial}{\partial\zeta} D^-_{\kappa+i/2})\Big(\frac{\vert m\vert +1}{2} -i (\kappa +i/2)\Big),\nonumber\\
\lambda^{1/2}_{-\kappa+i/2}&=&\frac{1}{2}\frac{\partial}{\partial\zeta}\sqrt{\zeta}D_{\kappa+i/2,m}^+ -2(\kappa+i/2) \tilde{c}^0_{\kappa -i/2} \nonumber\\&-&\frac{1}{2}\frac{\sqrt{\zeta}}{2}
\frac{\partial}{\partial\zeta} D^-_{\kappa+i/2} +\Big(\frac{\vert m\vert +1}{2} - i(\kappa-i/2)\Big) \Big(-i \tilde{c}^0_{\kappa +i/2} + \frac{1}{2}\frac{\sqrt{\zeta}}{2}
\frac{\partial}{\partial\zeta} D^-_{\kappa-i/2}\Big),\nonumber\\
\lambda^{-1/2}_{-\kappa-i/2}&=&\frac{1}{2}\frac{\partial}{\partial\zeta}\sqrt{\zeta}D_{\kappa-i/2,m}^+ -2(\kappa-i/2) \tilde{c}^0_{\kappa +i/2} \nonumber\\&-&\frac{1}{2}\frac{\sqrt{\zeta}}{2}
\frac{\partial}{\partial\zeta} D^-_{\kappa-i/2} +\Big(\frac{\vert m\vert +1}{2} + i(\kappa+i/2)\Big) \Big(+i \tilde{c}^0_{\kappa -i/2} + \frac{1}{2}\frac{\sqrt{\zeta}}{2}
\frac{\partial}{\partial\zeta} D^-_{\kappa+i/2}\Big),\nonumber\\
\lambda^{-3/2}_{-\kappa-3i/2} &=& (-i\tilde{c}^0_{\kappa +i/2} - \frac{\sqrt{\zeta}}{2}
\frac{\partial}{\partial\zeta} D^-_{\kappa-i/2})\Big(\frac{\vert m\vert +1}{2} +i (\kappa -i/2)\Big)\label{eq:lambda}.
\end{eqnarray}

For $\{\mathbf{E}_{\mathcal{E}},\mathbf{B}_{\mathcal{E}}\}$, and $m< 0$ the role of the functions $\pi_+$ and $\pi_-$ is interchanged, since the hypergeometric function involves the absolute value of $m$.

\begin{center}
{\bf Appendix C.} {Parabolic mirror boundary condition for $\zeta_0\gg \kappa$.}
\end{center}

Under standard conditions, the mirror typical lengths are much greater than the light wavelength of interest. In order to facilitate the numerical evaluation
of the values of $\kappa$ that guarantee that the condition Eq.~(\ref{eq:bc}) is satisfied, it is necessary to study the asymptotic behavior of the $\mathcal{W}_{\kappa,m}(\zeta_0)$ function.
We perform that analysis in two steps. First, we write the $V_{\kappa\pm i/2,m}(\zeta_0)\equiv V_\pm(\zeta_0)$ function in terms of Coulomb functions, Eq.~(\ref{eq:coul}). Then we use the asymptotic expressions for those 
functions \cite{abra}.
Let us define
\begin{equation}
\mathcal{F}_{\kappa,m}(\zeta_0) = \frac{F_{(\vert m +1\vert +1)/2}(\kappa,\zeta_0/2)}{F_{(\vert m -1\vert +1)/2}(\kappa,\zeta_0/2)}.
\end{equation} 
Using Eq.~(\ref{eq:v1}) and Eq.~(\ref{eq:v2}) it results that
\begin{equation}
\mathcal{F}_{\kappa,m}(\zeta_0) = -i\frac{(V_+ - V_-)\vert d_+\vert}{d_+ V_- + d_- V_+} = -i\frac{(\mathcal{W}_{\kappa,m}-1)\vert d_+\vert}{d_+ - d_-\mathcal{W}_{\kappa,m}}.
\end{equation}
So that,
\begin{equation}
\mathcal{W}_{\kappa,m} = -\frac{i{\mathcal{F}}d_+ + \vert d_+\vert}{i{\mathcal{F}}d_- + \vert d_-\vert},
\end{equation}
and the boundary condition 
\begin{equation}
\mathcal{W}^2_{\kappa,m}= -\frac{d_+}{d_-}, \label{eq:BcB}
\end{equation}
is equivalent to
\begin{equation}
\mathcal{F}^2_{\kappa,m}(\zeta_0)= 1.
\end{equation}
There is an infinite number of roots $\{\kappa_0\}$ of this equation. For $\zeta_0 >>1$ these roots become almost equidistant as $\kappa$ increases. 
In fact, for $\zeta_0 >> \kappa$, and $m>0$, the boundary condition can be written as
\begin{equation}
\frac{\cos^2 \theta_{\kappa,m}}{\tan^2 (\varphi_{\kappa,m})}\Big[ 1 - \tan^2\varphi_{\kappa,m} -2(\tan \theta_{\kappa,m})(\tan \varphi_{\kappa,m}) \Big]  \sim 0
\end{equation}
$$ \theta_{\kappa,m} ={\mathrm{arctan}}2\kappa/m;\quad\quad \varphi_{\kappa,m} = \zeta_0/2 - \kappa\log \zeta_0 - \frac{\vert m\vert}{2}\frac{\pi}{2} +
{\mathrm{arg}}\Gamma(md_+).$$
Some roots of Eq.~(\ref{eq:BcB}) are directly identified, 
\begin{equation}
\frac{2\ell +1}{2}\pi = \theta_{\kappa_\ell,m}, 
\end{equation}
with $\ell$ an integer number.

\end{document}